\documentclass[pdflatex,sn-mathphys-num]{sn-jnl}% Math and Physical Sciences Numbered Reference 
\usepackage{graphicx} % Required for inserting images
\usepackage{subcaption} % labeling for several plots in a figure
\usepackage[justification=raggedright, singlelinecheck=false]{caption}

\usepackage{amsmath}
\usepackage{amssymb}
\usepackage{braket}
\usepackage{hyperref}
\usepackage{xcolor}

\usepackage{placeins} % provides FloatBarrier command
\usepackage{comment}
\usepackage{tabularx}

\newcommand{\eps}{\ensuremath{\varepsilon}}
\newcommand{\CNOT}{\ensuremath{\mathrm{CNOT}}}
\newcommand{\CNOTp}{\ensuremath{\mathcal{CNOT}}}

\theoremstyle{thmstyleone}%
\theoremstyle{thmstyletwo}%
\newtheorem{remark}{Remark}%

\theoremstyle{thmstylethree}%

\begin{document}

\title{Real-time measurement error mitigation for one-way quantum computation}
\author[1,2]{\fnm{Tobias} \sur{Hartung}}\email{tobias.hartung@desy.de}
\equalcont{These authors contributed equally to this work.}
\author[3]{\fnm{Stephan} \sur{Schuster}}\email{stephan.schuster@fau.de}
\equalcont{These authors contributed equally to this work.}
\author[3]{\fnm{Joachim} \sur{von Zanthier}}\email{joachim.von.zanthier@fau.de}
\author*[4,5]{\fnm{Karl} \sur{Jansen}}\email{karl.jansen@desy.de}
\affil[1]{\orgdiv{Computing and Information Systems}, \orgname{Northeastern University - London}, \orgaddress{\street{Devon House, St Katharine Docks}, \city{London}, \postcode{E1W 1LP}, \country{United Kingdom}}}
\affil[2]{\orgdiv{Khoury College of Computer Sciences}, \orgname{Northeastern University}, \orgaddress{\street{\#202, West Village Residence Complex H, 440 Huntington Ave}, \city{Boston}, \postcode{02115}, \state{MA}, \country{USA}}}
\affil[3]{\orgdiv{Quantum Optics and Quantum Information Group}, \orgname{Friedrich-Alexander-Universität Erlangen-Nürnberg}, \orgaddress{\street{Staudtstr. 1}, \city{Erlangen}, \postcode{91058}, \country{Germany}}}
\affil[4]{\orgdiv{Computation-Based Science and Technology Research Center}, \orgname{The Cyprus Institute}, \orgaddress{\street{20 Kavafi Street}, \city{Nicosia}, \postcode{2121}, \country{Cyprus}}}
\affil[5]{\orgname{Deutsches Elektronen-Synchrotron DESY}, \orgaddress{\street{Platanenallee 6}, \city{Zeuthen}, \postcode{15738}, \country{Germany}}}

\abstract{
    We propose a quantum error mitigation scheme for single-qubit measurement errors, particularly suited for one-way quantum computation. Contrary to well established error mitigation methods for circuit-based quantum computation that require the circuits to be run several times, our method is capable of mitigating measurement errors in real-time, during the processing measurements of the one-way computation. For that, an ancillary qubit register is entangled with the to-be-measured qubit and additionally measured afterwards. By using a voting protocol on all measurement outcomes, occurring measurement errors can be mitigated in real-time while the one-way computation continues. 
    We provide an analytical expression for the probability to detect a measurement error depending on the error rate and the number of ancilla qubits. From this, we derive an estimate of the ancilla register size for a given measurement error rate and a required success probability to detect a measurement error. Additionally, we also consider the CNOT gate error in our mitigation method and investigate how this influences the probability to detect a measurement error. Finally, we show in proof-of-principle simulations and in experiments executed on IBM quantum hardware that our method is capable of reducing the measurement errors significantly in a one-way quantum computation with only a small number of ancilla qubits.
}

\maketitle

\section{Introduction}
\label{sect:introduction}
During the current era of noisy intermediate-scale quantum devices (NISQ devices), harnessing the full potential of the available quantum processors is a necessary but demanding task for all quantum computation applications. 
While the quantum circuit model is the most well-known and most used quantum computational model on current NISQ devices, also several models of measurement-based quantum computation (MBQC) exist~\cite{Raussendorf2001, Raussendorf2003, Briegel_2009, Bartolucci2021} and have received increasing interest in the search for scalable quantum computation~\cite{Ferguson2021, Marqversen_2023, Schuster_2024, Qin_2024, Brandhofer2024, Romero2024}. 

In MBQC, the computation is performed via measurements and local operations on a special class of entangled resource states, called graph states~\cite{Raussendorf2001, Raussendorf2003, Briegel_2009, Bartolucci2021}. Those MBQC models are computationally equivalent to the circuit-based model, while their fundamental concepts and technical requirements can differ significantly~\cite{Briegel_2009, Bartolucci2021}, since the MBQC paradigm trades larger memory for less complex quantum operations. The abstract MBQC paradigm consists of three steps: entangle an ensemble of qubits, measure a set of ancillae, and correct the output based on the measurement outcomes. The correct identification of required corrective measures is therefore paramount. Hence, the focus of the work presented here is the real-time verification of such corrective measures in a way that is compatible with a specific MBQC model called one-way quantum computing (OWQC), a measurement-based model proposed by Raussendorf and Briegel~\cite{Raussendorf2001, Raussendorf2003, Briegel_2009}. 

Recently, it has been shown that using elements from MBQC, such as mid-circuit measurements and feedforward quantum operations, in quantum circuits (dynamic circuits) can boost their performance and yield an advantage over unitary-only circuits~\cite{Pino_2021, Yirka_2021, Chertkov_2022, Chertkov_2023, DeCross_2023, Chan_2024, B_umer_2024}. A protocol that allows for verification of correct feedforward operations is thus applicable to dynamic circuits as well.

On the other hand, the variational quantum eigensolver, an extensively studied protocol for quantum circuits~\cite{Kokail2019, Nannicini2019, BravoPrieto2020, Funcke2023, Chan_2024, Sobhani2024, Schwaegerl2024, Guo2024, Chai2024, DiMeglio2024}, has recently been adapted for MBQC~\cite{Ferguson2021,  Marqversen_2023, Schuster_2024, Qin_2024}. 
However, the significant noise level in current NISQ devices makes error mitigation a necessary step in every quantum computation. As we will discuss in Section~\ref{subsect:postprocessing}, the misidentification of corrective operations has a particularly severe impact. The implementation of a verification method is therefore crucial for existing error mitigation techniques that rely on errors acting perturbatively to perform well.

Several methods to mitigate different types of errors are well established for unitary-only quantum circuits~\cite{Temme_2017, Giurgica_Tiron_2020, Bravyi_2021, Nation_2021, Berg_2022, Funcke_2022, van_den_Berg_2023, Funcke2023, Kurosawa2024}. For dynamic circuits and MBQC new methods to mitigate errors in final expectation values obtained from the circuit output state have been proposed~\cite{Gupta_2023, Koh_2024}. There has, however, been slow progress regarding the direct mitigation of measurement errors in the processing measurements and subsequent identification of corrective operations in MBQC. Using established mitigation methods of quantum circuits for MBQC has the major drawback that calibration runs need to be executed to learn the measurement error rates before the computation and errors can then only be compensated during post-processing~\cite{Bravyi_2021, Nation_2021, Berg_2022, Funcke_2022}. Since the computation in MBQC proceeds via the qubit measurements, a mitigation method that can be applied in real-time, identifying not only the state of the measured qubit but more importantly directly inferring the state of the to be corrected qubit, is required. 

In this paper, we implement a method that uses additional verification qubits for each single-qubit measurement and a voting protocol to determine the correct state of the qubit still remaining in a OWQC. This voting protocol approach is well-established in classical computing and some gate-based quantum computing protocols. However, the main focus of this work is compatibility with OWQC compute steps and the correct identification of the corrective operations. Instead of the correct identification of the measured ancilla state because projection/backaction errors may cause the measured ancilla state to not coincide with the state indicating the corrective operation that needs to be performed. In contrast to error correcting codes, which have severe overheads given the larger memory requirements in OWQC, we require only a constant number of verification qubits depending on the measurement error rates, by re-using the verification qubits after each measurement. 

We note that methods based on ancillary qubits have already been proposed to boost the performance of photonic Bell measurements~\cite{Grice2011, Ewert_2014, Bayerbach2023}, followed by experimental realizations~\cite{Bayerbach2023, Hauser2024}. However, those methods differ from our method in the qubit state preparations before the measurements, using specifically tailored preparation schemes for photonic Bell measurements~\cite{Grice2011, Ewert_2014, Bayerbach2023, Hauser2024}. Nevertheless, they still provide a valuable mitigation method for fusion-based quantum computation, which uses multi-qubit measurements instead of single-qubit measurements for the computation~\cite{Bartolucci2021,Hauser2024}.

The paper is organized as follows. First, in Section~\ref{sect:mitigation}, we discuss post-processing-based mitigation techniques, their limitations, and why we believe a voting protocol is preferable for the identification of corrective operations. We then introduce the model of measurement errors we are going to use in this work in Section~\ref{sect:modelling_readout_and_projection_errs}. We make a clear distinction between errors that occur through the readout apparatus after projecting the qubit during the measurement (readout errors) and the actual projection error that occurs during the projection since this is the foundational difference between correctly identifying the state of the measured ancilla and the corrective operation required for the computation to proceed. In Section~\ref{sect:compensating_readout_errs}, we discuss the compensation of readout errors via a voting protocol. Afterwards, in Section~\ref{sect:proj_error_mitigation}, we discuss a method to compensate projection errors using a register of verification qubits for each measurement in the OWQC. We additionally consider the $\CNOT$ error within this method in Section~\ref{sect:cnot_errs} and present results where we applied our method to the elementary compute step of every OWQC in Section~\ref{sect:simulation_res}. The results were obtained via numerical simulations and additionally through execution on IBM quantum hardware. Finally, we make some concluding remarks in Section~\ref{sect:conclusion}.

\section{Mitigation techniques}
\label{sect:mitigation}

The primary objective of this work is the application of mitigation protocols for OWQC. In this section, we will therefore review the two main mitigation approaches -- post-processing and voting protocols -- with regards to their suitability for OWQC.

\subsection{Post-processing}
\label{subsect:postprocessing}

The most common approaches to error mitigation in quantum computing are statistical post-processing schemes~\cite{Funcke_2022,Berg:2022ugn,PhysRevX.7.021050,PhysRevLett.119.180509,Kandala2018ErrorME,PRXQuantum.3.010345,PhysRevX.11.041036,Cai2021quantumerror,PhysRevLett.129.020502,9226505,Saxena:2024ria,Filippov:2024yhx}. In these frameworks, a quantum computation is repeated to obtain an outcome distribution. For example, one might use $S$ shots (repetitions) and observes a qubit in the state $\ket0$ with frequency $f_0$ and in $\ket 1$ with frequency $f_1$. Using machine calibration data or specific modeling techniques, we know that the observed frequency vector $f=(f_0,f_1)^T$ can be modeled via a perturbation matrix $P$ from the ``true'' (error-free) distribution vector $F$ as $f=PF$. Thus, instead of using the observed frequency vector $f$, expectations of observables are computed with $P^{-1}f$. Of course, there exist more elaborate techniques where suitable additional noise is introduced to the circuit to allow for mitigation of expectation values either directly or via additional post-processing such as an extrapolation step in the zero noise extrapolation.

The foundational premise of these type of mitigation approaches is therefore that a circuit is executed multiple times for expectation values of observables to be better approximated. However, this is not what is required for OWQC. A compute step in OWQC measures a qubit which affects an operation on another qubit. Based on the measurement outcome, one of two operations is performed, and a corrective operation may be required to ensure that the affected qubit has the same operation performed on it independent of the measurement outcome of the measured qubit. This implies that there are two fundamental obstacles to post-processing approaches within the OWQC compute step. First and foremost, repeat evaluations of the circuit are not possible for a single compute step as the computation does not end with the measurement step. This approach is perfectly viable for the final outcomes, but not for mitigating individual compute steps. Second, the distribution of outcomes that post-processing aims to correct is irrelevant for the compute step. The measurement outcome is irrelevant per se. The ability to correctly identify the operation performed on the affected qubit is what matters.

For example, suppose that we want to execute an operation $U$ on the affected qubit, but the measured qubit indicates that an operation $U'=C^\dagger U$ has been performed. We then correct the state of the affected qubit by applying the correction $C$ to it. In the correction-measurement-entanglement (CME) standard pattern~\cite{10.1145/1219092.1219096}, this operation $C$ is a product of Pauli $X$ and $Z$ operations. Suppose the current computational state is $\ket{\psi_0}$ and the measured qubit is misidentified. Then, a wrong correction $C$ is applied to $\ket{\psi_0}$. Thus, the computation continues with an incorrect state $\ket{\tilde\psi}$ while the correct state should have been $\ket{\psi}$ with $\ket{\tilde\psi}=P\ket\psi$ where $P$ is a (multi-qubit) Pauli matrix. Hence, the Hilbert space can be split into two subspaces of equal dimension corresponding to the eigenspaces with respect to $\pm1$ of $P$. We can then decompose $\ket\psi=\alpha\ket++\beta\ket-$ with respect to those eigenspaces as well and obtain $\ket{\tilde\psi}=\alpha\ket+-\beta\ket-$. The angle between $\ket\psi$ and $\ket{\tilde\psi}$ is therefore $\theta=\arccos(|\alpha|^2-|\beta|^2)=\arccos(2|\alpha|^2-1)$. This is equivalent to a fidelity of $|2|\alpha|^2-1|$. Since unitary operations are angle-preserving, this angle between the computational state and the true state it should be in remains unchanged until and unless another measurement is misidentified.

Assuming a $Q$-qubit state $\ket{\psi}$ chosen uniformly at random, we can model $|\alpha|^2$ as a quotient $\frac{A}{A+B}$ where $A$ and $B$ are sums of $2^{Q}$ independent standard-normal random variables, i.e., $A$ and $B$ are $\chi^2_{2^Q}$-distributed. Thus, $|\alpha|^2$ has a beta-distribution $B(2^{Q-1},2^{Q-1})$, i.e., it has an expected value of $\frac12$. Furthermore, the expected angle is $\mathbb{E}(\theta)=\frac{\Gamma(2^{Q+1})}{\Gamma(2^Q)^2}\int_0^1\arccos(2x-1)x^{2^Q-1}(1-x)^{2^Q-1}dx=\frac{\pi}{2}$.
For the expected fidelity, we note that $2|\alpha|^2-1$ has mean $0$ and standard deviation $\frac{1}{2\sqrt{2^{2Q}+1}}$ which implies that for large numbers of qubits $Q$, the expected fidelity after a single misidentification is very close to zero with very high confidence.

This fundamentally forces an attempt at post-processing-based mitigation techniques to be implemented in such a way that statistical evaluation becomes possible. Theoretically, this can be achieved using deferred measurements. However, this effectively replaces the measurement feedforward operation of OWQC with a de facto gate-based computation. While this is in principle possible, it prevents reusability of measured qubits, introduces more complex operations, and thereby removes the advantages of OWQC.

\subsection{Voting protocols}
\label{subsect:voting}

The discussion in Subsection~\ref{subsect:postprocessing} above indicates that standard post-processing approaches to quantum computation may be relevant in mitigating errors equivalent to errors arising in gate-based computation, but they cannot address the misidentification error which is crucial for OWQC. As seen above, given a single misidentification, the resulting state is expected to be orthogonal to the state the machine should be in, and for large numbers of qubits, a single misidentification leads to a complete loss of fidelity. For OWQC it is therefore paramount to correctly identify which operation needs to be performed on the computational state.

The most common approach to achieve such a goal is to implement majority signal voting. Majority signal voting has a long history in various fields of engineering ranging from early communications using increased power levels and Hamming codes to improve reliability, to modern sensor technology where data from multiple sensors is averaged and sensors whose data is too far from the average get labeled as unreliable and excluded~\cite{Avila_Rodriguez_2011, Orr_2002, Pratt_2005, Spilker_1998, Hamming_1950, Moon_2005}.

\begin{figure}
    \centering
    \includegraphics[width=\textwidth]{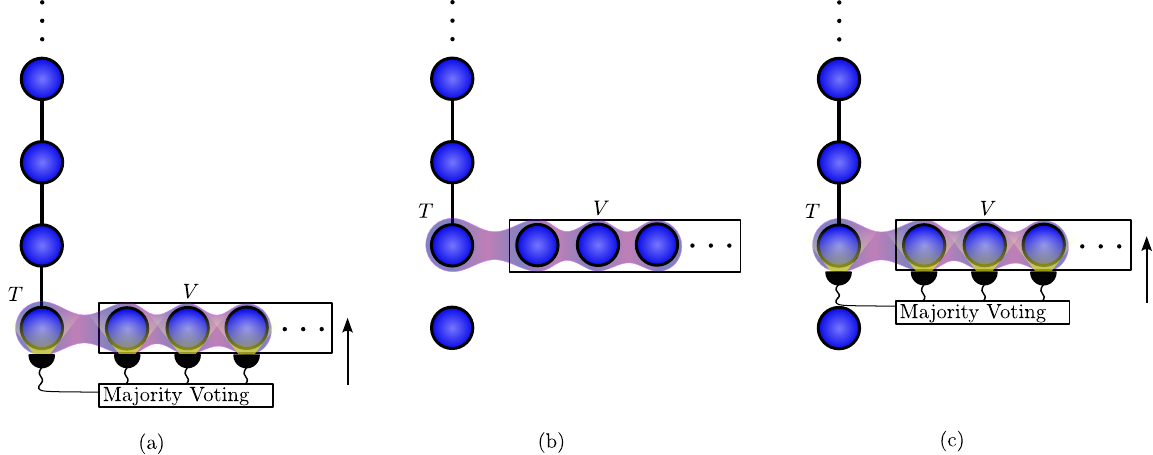}
    \caption{Schematic implementation of the verified measurement protocol with qubit reuse . The linear graph represents a OWQC, where the target qubit $T$ should be measured next. The verification register $V$ is shown here with 3 qubits. Panel (a) - (c) show a full cycle during the OWQC, consisting of (a) a verified measurement on qubit $T$ and register $V$, (b) the re-entangling of $V$ with the next target qubit and (c) the next measurement.}
    \label{fig:owqc_measurements_mitigation_scheme}
\end{figure}

Error correcting codes in quantum computing naturally fall into this approach~\cite{Shor_1995, Steane_1996, Kitaev_1997, Bravyi_1998, Steane_1996a, DiVincenzo_1996}. However, using multiple physical qubits to encode a single logical qubit incurs significant overheads. This is particularly prohibitive for OWQC compared to gate-based models, since every measurement-based operation requires a qubit to be measured. Thus, the number of logical qubits required scales with the number of operations performed. For example, performing an Euler rotation on a single qubit requires $4$ additional qubits in OWQC. Thus, using $q$ physical qubits per logical qubit and performing $k$ Euler rotations on that physical qubit requires $q$ physical qubits in the gate-based model but $(4k+1)q$ physical qubits in OWQC. This renders error correcting codes near-term infeasible for OWQC. On the other hand, it should be noted that once a qubit has been measured, it is available for further calculation. Thus, we aim to implement a voting protocol that reuses a verification register at different times as shown in Fig.~\ref{fig:owqc_measurements_mitigation_scheme}. 

The implementation of such a ``verified measurement'' through the use of a verification register requires the encoding of single qubits into entangled multi-qubit states immediately prior to the measurement and readout procedure (cf. Fig.~\ref{fig:owqc_measurements_mitigation_scheme}b). Within the context of gate-based quantum computing, various such implementation approaches have been considered~\cite{PhysRevA.105.012419,Guenther_2022,Kim_2025,kim2025protocolpurifyingnoisypreparation,8tph-mc2p,van_den_Berg_2023_2}. The approaches considered in~\cite{PhysRevA.105.012419,Guenther_2022,Kim_2025,kim2025protocolpurifyingnoisypreparation,8tph-mc2p,van_den_Berg_2023_2} are in principle also applicable to OWQC and, if applied to OWQC compute steps, their results remain valid.

However, there are also substantial differences between those voting methods, the commonly used post-processing methods, and our own framework. For the most relevant approaches, we summarized the major differences in Tab.~\ref{tab:comparison}, covering if the method requires calibration, or post-processing, on what exactly the mitigation acts on and if this method is applicable to measurement outcomes in real-time.
\begin{table}[t]
\begin{tabularx}{\textwidth}{Xcccc}
\hline
Approach & Calib. & Post- & Acts on & Real- \\
 & required & processing & & time \\
\hline
Post-processing mitigation~\cite{Bravyi_2021,Nation_2021,Funcke_2022,Berg_2022,Koh_2024}
  & yes & yes & statistics$^{a}$ & no \\
Measurement purification~\cite{Kim_2025,kim2025protocolpurifyingnoisypreparation}
  & no & no & single shots & no \\
Ancilla-encoded readout~\cite{Guenther_2022,PhysRevA.105.012419}
  & no & no & single shots & yes \\
Verified measurements (This work)
  & no & no & feedforward operation & yes \\
\hline
\end{tabularx}
{\footnotesize
$^{a}$ Ref.~\cite{Koh_2024} addresses mid-circuit measurements that drive feedforward
operations, but mitigates only the ensemble expectation values; the operation applied in an
individual run remains uncorrected.
}
\caption{Comparison of representative measurement-error mitigation approaches with the
framework proposed in this work. ``Calibration'' refers to device characterization that must be
performed before the computation, ``Post-processing'' refers to a required post-processing after the measurements to mitigate the results, ``acts on'' denotes the quantity that is actually corrected and
``real-time'' refers to the ability to correct a single compute step while the computation
continues.}
\label{tab:comparison}
\end{table}
Tab.~\ref{tab:comparison} shows that the approaches of~\cite{PhysRevA.105.012419,Guenther_2022} are very similar to our work. They propose a similar method to our verified measurement approach for computational basis measurements in gate-based quantum computing. There is, however, a gap between measurement verification in OWQC and gate-based computing in the fact that traditional gate-based computing always measures in the computational basis at the end of a quantum circuit\footnote{Both works do not consider any mid-circuit measurements or feedforward operations~\cite{PhysRevA.105.012419,Guenther_2022}.}, while in OWQC the operations performed on the device state are determined by the choice of measurement basis, and the resulting (error-free) measurement outcomes determine the feedforward operations. The work presented in the following therefore primarily aims at bridging this gap, to make measurement verification directly targeting the feedforward operations and thereby applicable to OWQC.

\section{Modelling readout and projection errors for single-qubit measurements}
\label{sect:modelling_readout_and_projection_errs}
Considering a single-qubit measurement in an arbitrary basis $\{\ket{e_1}, \ket{e_2}\}$, we can in general identify two kinds of measurement errors that can occur:
\begin{itemize}
    \item Given that the qubit is in one of the measurement basis states $\ket{b}\in\{\ket{e_1},\ket{e_2}\}$, a projection error may project this state during the measurement into the orthogonal basis state $\ket{\neg b}$ with a probability $p_{\neg b, b}$.
    \item Given that the qubit was projected into $\ket{b}$, the readout apparatus may falsely record the qubit to be in state $\ket{\neg b}$. Which we call a readout error, occurring with probability $r_{\neg b, b}$.
\end{itemize}

While it is not common to separate projection and readout error in the quantum computing literature, it is a common separation in other fields such as nuclear magnetic resonance imaging where it is usually discussed within the context of sensor backaction, non-destructive readouts, or non-demolition measurements~\cite{Holzgrafe:2018cap,doi:10.1126/science.1189075,Fan:22,doi:10.1126/sciadv.aaw6664,Rosskopf2016AQS,PhysRevLett.99.120502,Pfender:2016qke}. It is also useful and important within the context of OWQC because we have qubit pairs containing an ``upstream'' and a ``downstream'' qubit, and we make a computational decision based on the measurement outcome of the upstream qubit. The distinction becomes important for OWQC precisely because we infer the state of the downstream qubit from the measurement outcome of the upstream qubit. As such, the measured qubit acts as a witness about the state of another qubit instead of giving information about its own state, which is the more canonical usage of measurement outcomes. In the latter case, the distinction becomes less relevant.

 For simplicity, let us assume that the qubit pair is in a Bell state $\frac{\ket{00}+\ket{11}}{\sqrt{2}}$. Measuring the upstream qubit should collapse the state into either $\ket{00}$ or $\ket{11}$ and reading the state of the upstream qubit tells us the state of the downstream qubit. Without loss of generality, let the downstream qubit be in the $\ket0$ state. A readout error now may occur when the upstream qubit is also in the $\ket0$ state but our measurement device records a $1$ instead of the correct $0$. This is clearly problematic as we have misidentified the state of the downstream qubit and decide to perform an incorrect operation on it.

A projection error on the other hand occurs when the state of the upstream qubit is incorrectly projected due to the energy introduced to the system through the measurement process, i.e., although the joint state collapsed into the $\ket{00}$ state, the energy transferred onto the upstream qubit during measurement causes it to flip and end up in the $\ket1$ state instead. That is, a projection error leads to the ``impossible'' outcome $\ket{10}$ for the qubit pair. Even under correct readout, this too would lead to a misidentification of the downstream qubit and an erroneous computation due to the resulting wrong computational decision.

 In practice, the two errors have very different behavior as can be seen considering a qubit in the state $\ket0$ and performing repeated projective measurements on it. A readout error maintains the underlying state $\ket0$, i.e., a small readout error chance would lead to a measurement outcome sequence similar to $0,\ldots,0,1,0,\ldots,0$. Meanwhile a projection error changes the underlying state. A small projection error chance in repeated measurements therefore leads to a measurement outcome sequence similar to $0,\ldots,0,1,\ldots,1$.

As such, separating the two errors in standard quantum computing setups is not necessary because both have the effect of shifting the measurement distribution of repeated runs in the same way and can therefore be mitigated via post-processing in the same way. However, in OWQC we need to correctly identify the state of each individual downstream qubit, thus preventing post-processing-based mitigation approaches. Importantly, hardware design approaches that allow for separation of projection and readout error (e.g.,~\cite{Holzgrafe:2018cap,doi:10.1126/science.1189075,Fan:22,doi:10.1126/sciadv.aaw6664}) are beneficial for reducing error mitigation overheads in OWQC.

To further clarify the separation of projection and readout error, suppose that a qubit is in the state $\ket{\psi}=\alpha\ket{0}+\beta\ket{1}$. We will obtain the following probabilities for a single-qubit computational basis measurement if no errors are present:
\begin{center}
\begin{tabular}{|c||c|c|}
\hline
   projected  & recorded $0$ & recorded $1$ \\
   \hline
    $\ket{0}$ & $\lvert \alpha \lvert^2$ & $0$ \\
    \hline
    $\ket{1}$ & $0$ & $\lvert \beta \lvert^2$ \\
    \hline
\end{tabular}
\end{center}
This table changes as follows if we consider now projection errors to occur.
\begin{center}
\begin{tabular}{|c||c|c|}
\hline
   projected  & recorded $0$ & recorded $1$ \\
   \hline
    $\ket{0}$ & $p_{0,0}\lvert \alpha \lvert^2+p_{0,1}\lvert\beta\lvert^2$ & $0$ \\
    \hline
    $\ket{1}$ & $0$ & $p_{1, 1}\lvert \beta \lvert^2+p_{1, 0}\lvert\alpha\lvert^2$ \\
    \hline
\end{tabular}
\end{center}
Here, $p_{0, 0}$ (and $p_{1,1}$) represent the success probabilities that the qubit which is initially in state $\ket{0}$ (and $\ket{1}$) is also projected into state $\ket{0}$ (and $\ket{1}$). If the outcomes of these measurements can additionally fail to be recorded correctly with probabilities $r_{0,1}$ and $r_{1,0}$, we arrive at the following table.
\begin{center}
\begin{tabular}{|c||c|c|}
\hline
   projected  & recorded $0$ & recorded $1$ \\
   \hline
    $\ket{0}$ & $r_{0,0}(p_{0,0}\lvert \alpha \lvert^2+p_{0,1}\lvert\beta\lvert^2)$ & $r_{1,0}(p_{0,0}\lvert \alpha \lvert^2+p_{0,1}\lvert\beta\lvert^2)$ \\
    \hline
    $\ket{1}$ & $r_{0,1} (p_{1, 1}\lvert \beta \lvert^2+p_{1, 0}\lvert\alpha\lvert^2)$  & $r_{1,1} (p_{1, 1}\lvert \beta \lvert^2+p_{1, 0}\lvert\alpha\lvert^2)$ \\
    \hline
\end{tabular}
\end{center}
Similarly, $r_{0,0}$ and $r_{1,1}$ represent the success probabilities that the state is recorded to be in the same state that it was also projected to. 

For this work, we will assume that all considered error probabilities $p_{\neg b,b}$, $r_{\neg b,b}$ are sufficiently small such that measurements are statistically different from fair coin flips.

\section{Compensating readout errors}
\label{sect:compensating_readout_errs}
For the compensation of the readout errors, we will assume that the qubit has already been projected into its final state and that it can be read out without any further projections\footnote{The mitigation of projection errors will be discussed afterwards in the next section.}. Separating readout from projective measurements can be achieved with hardware designs that limit measurement backactions such as proposed in~\cite{Holzgrafe:2018cap} or via non-demolition schemes~\cite{doi:10.1126/science.1189075}. In this case the remaining probabilities will be $r_{b', b}$, that a qubit in state $\ket{b}$ is identified as being in state $\ket{b'}$.

$N$ successive readouts of a qubit in state $\ket b$ yield a binomial outcome distribution with probability $P(N,k)$ of $k$ correctly recorded results, 
\begin{equation}
\label{eq:binomial_readout_dist}
  P(N,k)=\binom{N}{k}r_{b,b}^kr_{\neg b, b}^{N-k},
\end{equation}
if the readout errors are statistically independent and identically distributed.
Assuming $k$ correct and $N-k$ incorrect readouts, we employ a voting protocol to decide which readout is correct. In other words, if there are $n$ readouts of a qubit in the state $\ket0$ and $N-n$ readouts in the state $\ket 1$, then we decide that the qubit is in state $\ket0$ if $n> N-n$. In order to have a clearly defined result from voting, it is therefore necessary that $N$ is an odd number.

\begin{figure}
    \centering
    \includegraphics[width=0.9\textwidth]{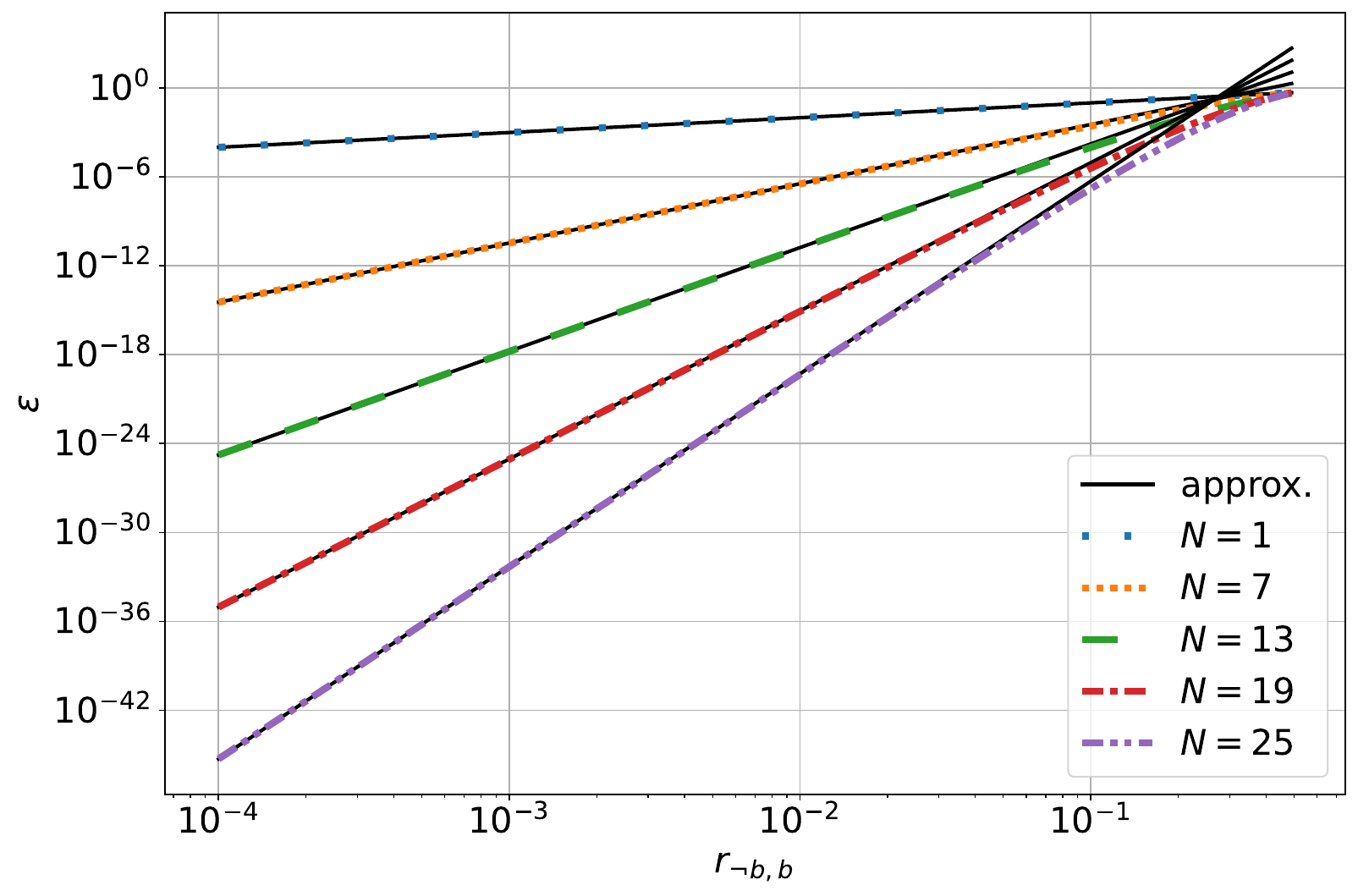}
    \caption{Probability of misidentification $\eps$ using a voting protocol of $N$ readouts from Eq.~\eqref{eq:misident_readout_err} in dependence of the (individual) readout error $r_{\neg b, b}\in[10^{-4},0.5]$. The solid black lines show the polynomial approximation of $\eps$ as shown in Eq.~\eqref{eq:misident_readout_err_poly_approx}.}
    \label{fig:misident_readout_err}
\end{figure}

The probability of misidentification given $N$ successive readouts is then given by the probability of having at most $\lfloor N/2\rfloor$ correct readouts
\begin{equation}
\label{eq:misident_readout_err}
\begin{aligned}
  \eps :=& \mathrm{prob}(\mathrm{misidentification}) \\
  =&\sum_{k=0}^{\lfloor N/2\rfloor}{\binom{N}{k}}r_{b,b}^{k}r_{\neg b, b}^{N-k}\\
  =&I_{r_{\neg b, b}}(N-\lfloor N/2\rfloor,1+\lfloor N/2\rfloor)
\end{aligned}
\end{equation}
where $I_x(a,b)=\frac{\Gamma(a+b)}{\Gamma(a)\Gamma(b)}\int_0^xt^{a-1}(1-t)^{b-1}dt$ denotes the regularized incomplete beta function. Fig.~\ref{fig:misident_readout_err} shows this misidentification probability for different number of readouts $N$ in dependence of the readout error $r_{\neg b, b}$. 

Fig.~\ref{fig:misident_readout_err} shows the potential improvements of implementing this voting protocol over a large range of (base) readout errors $r_{\neg b,b}$. In particular, for realistic error rates of no more than $5\%$ and $N\le25$, the polynomial approximation \begin{equation}
\label{eq:beta_poly_approx}
\begin{aligned}
    I_x(a,b)=&\frac{\Gamma(a+b)}{\Gamma(a)\Gamma(b)}\int_0^xt^{a-1}(1-t)^{b-1}dt\\
    =&\frac{\Gamma(a+b)}{\Gamma(a)\Gamma(b)}\int_0^xt^{a-1}\sum_{j=0}^{b-1}{\binom{b-1}{j}}(-t)^jdt\\
    =&\frac{\Gamma(a+b)}{\Gamma(a)\Gamma(b)}\sum_{j=0}^{b-1}{\binom{b-1}{j}}(-1)^j\frac{x^{a+j}}{a+j}\\
    =&\frac{\Gamma(a+b)}{a\Gamma(a)\Gamma(b)}x^a+\mathcal{O}(x^{a+1})
\end{aligned}
\end{equation}
seems to be useful for at least order of magnitude estimates. 

Of course, in general we are interested in predicting the value of $N$ given an expected probability of misidentification $\eps$. 
This dependence of $N$ on $\varepsilon$ within the polynomial approximation is shown in Eq.~\eqref{eq:N_eps} and Fig.~\ref{fig:N_by_err_a}:
\begin{figure}
    \centering
    \begin{subfigure}{0.45\textwidth}
        \centering
        \includegraphics[width=\textwidth]{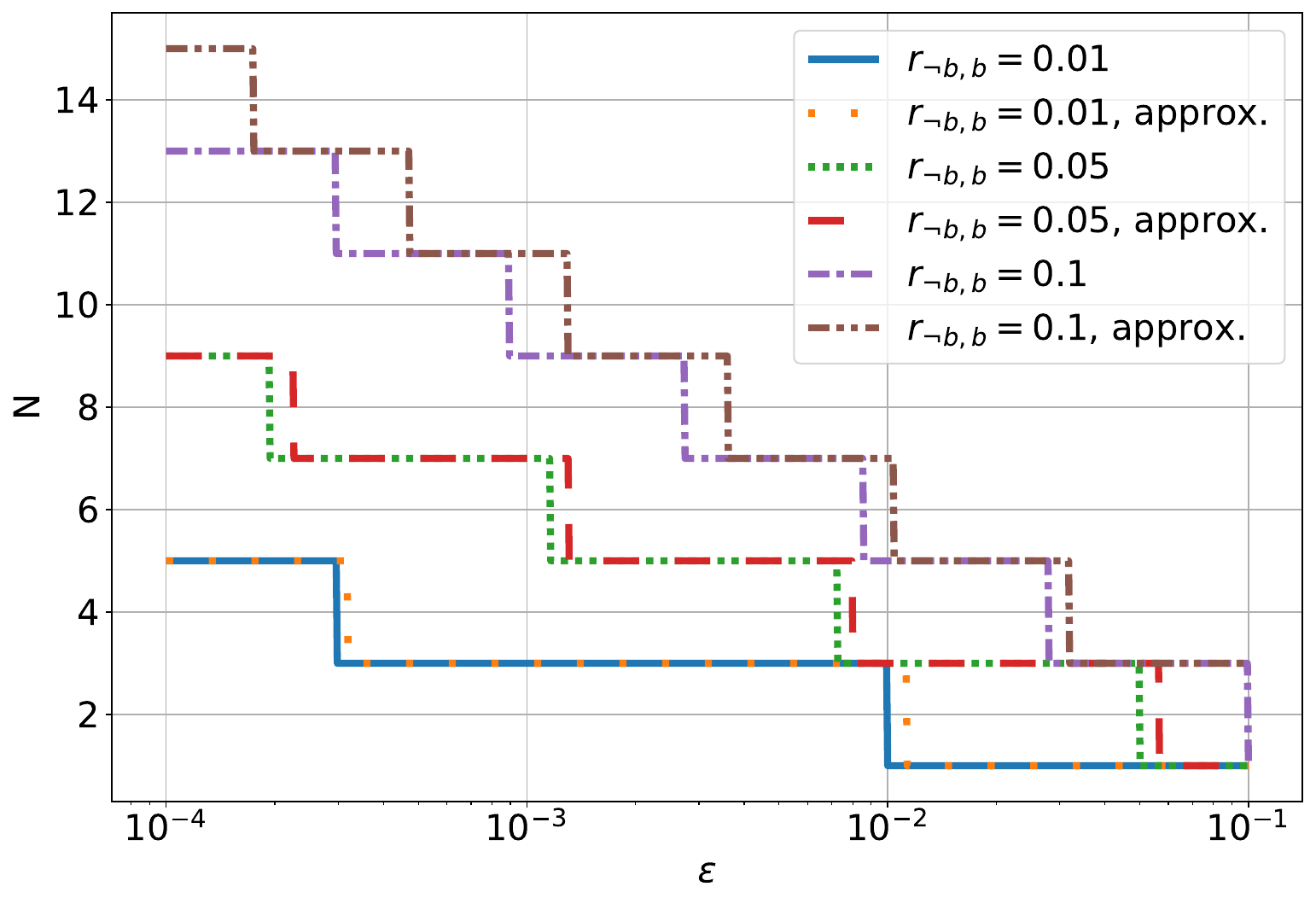} 
        \captionsetup{justification=centering}
        \caption{}
        \label{fig:N_by_err_a}
    \end{subfigure}
    \hfill
    \begin{subfigure}{0.45\textwidth}
        \centering
        \includegraphics[width=\textwidth]{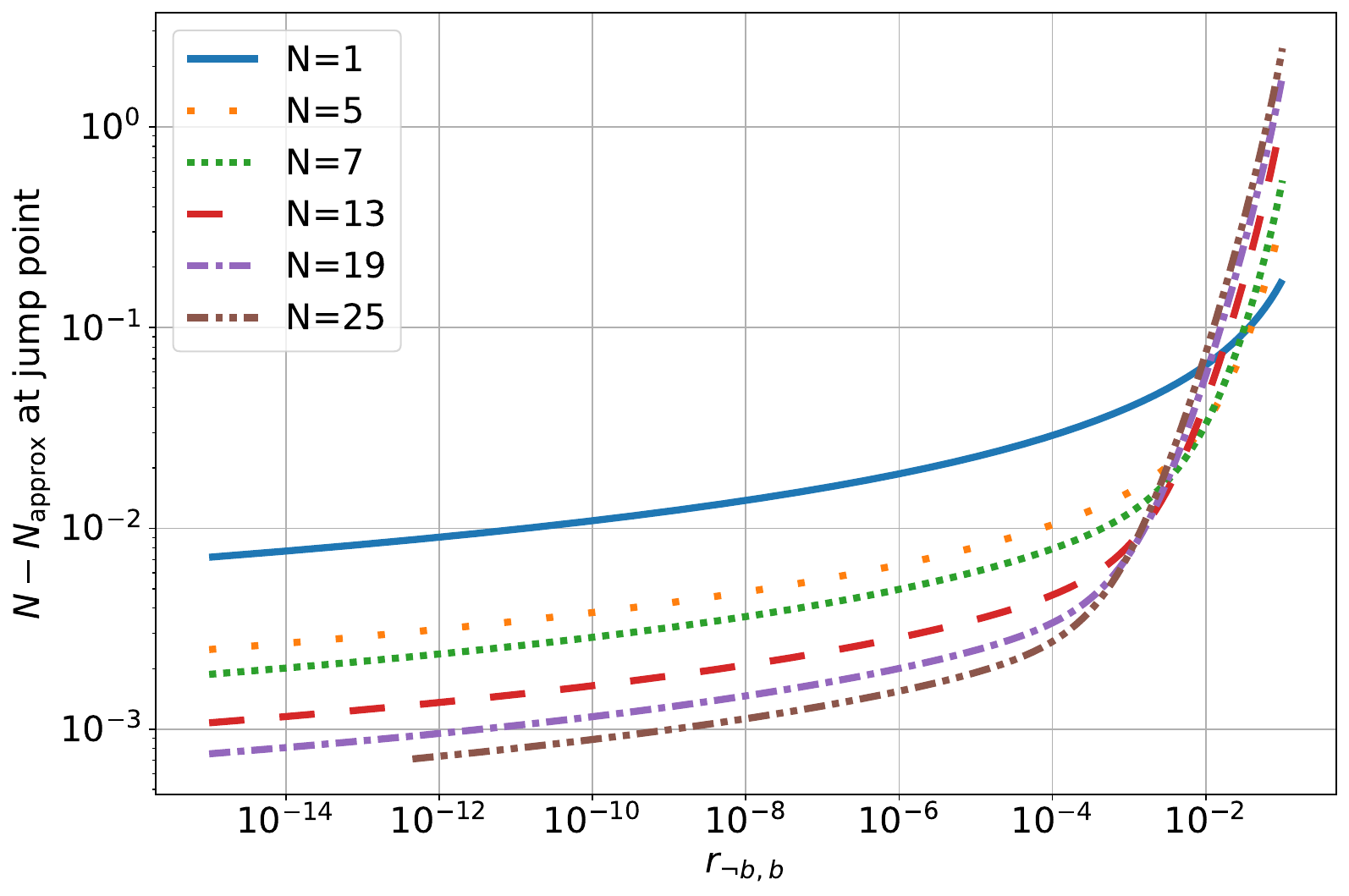} 
        \captionsetup{justification=centering}
        \caption{}
        \label{fig:N_by_err_b}
    \end{subfigure}
    \caption{Panel (a): Minimum number of readouts $N$ required to achieve a given probability of misidentification $\varepsilon$ obtained by solving Eq.~\eqref{eq:misident_readout_err} for $N$ using multiple (individual) readout error rates $r_{\neg b, b}\in[0.01,0.1]$. The approximations are calculated via Eq.~\eqref{eq:N_eps} and rounded to the next higher odd number. Panel (b): Difference in $N$ obtained by using Eq.~\eqref{eq:misident_readout_err} and Eq.~\eqref{eq:N_eps} in dependence of $r_{\neg b, b}$ at the jump positions of $N(\eps)$ (cf. panel (a)).}
    \label{fig:N_by_err}
\end{figure}

\begin{equation}
\label{eq:N_eps}
    N(\eps) = \frac{-W(-\frac{\ln(4r_{\neg b, b})}{2\pi\eps^2})}{\ln(4r_{\neg b, b})}- 1,
\end{equation}
where $W$ denotes the Lambert $W$ function. The derivation of Eq.~\eqref{eq:N_eps} is presented in the appendix. Since Eq.~\eqref{eq:N_eps} is a continuous function, we have to round its value to the next higher odd number,
\begin{equation*}
    N(\eps) \mapsto \begin{cases}
        \left\lceil N(\eps)\right\rceil\phantom{+1} \quad\text{ if } \left\lceil N(\eps)\right\rceil \text{ is odd}\\
        \left\lceil N(\eps)\right\rceil+1 \quad\text{else }
    \end{cases}.
\end{equation*}
As can be seen from Fig.~\ref{fig:N_by_err_b}, the jumps of $N$ from one odd number to the next occur at higher $\eps$ for the approximation in Eq.~\eqref{eq:N_eps} than for the iterative calculation. This means that we predict slightly higher $N$ with our approximation than actually required, within a certain region of the actual jump positions. This region gets larger for larger error rates $r_{\neg b,b}$. This may be caused by the increasing deviation of the polynomial approximation from the actual regularized incomplete beta function for higher measurement errors.

\section{Mitigating projection errors in a OWQC with error-free CNOT operations}
\label{sect:proj_error_mitigation}

As shown in Section~\ref{sect:compensating_readout_errs}, readout errors can be mitigated effectively via repeated readouts, if the readout procedure is separated from the projective measurements. Due to the fact that projection errors can change the underlying state of the qubit on repeated measurements, such a voting protocol based on measurement repetition is not possible for the projection errors. Thus, the voting protocol needs to be implemented through a larger verification qubit register and each qubit is only projected once. Hence, if no separation of readout and projection is possible, then the readout error $r_{\neg b,b}$ can be absorbed into the projection error $p_{\neg b,b}$ forming a combined measurement error. Such a combined error can be treated similarly to a pure projection error in our mitigation method. The implementation of a verification register to mitigate projection or combined measurement errors via majority voting will be discussed in the following.

On the lowest level of a one-way quantum computation (OWQC), we are working with graph states of two qubits which are in the state
\begin{equation}
\label{eq:HRz_owqc}
  \ket{\psi}_{TC}=CZ\ket{++}=\ket\alpha\otimes HR^z(\alpha)\ket++\ket{-\alpha}\otimes XHR^z(\alpha)\ket+.
\end{equation}
The qubit in the state $\ket{\pm\alpha}$ is the target $T$ for the projective measurement in the $\ket{\pm\alpha}$-basis and the second qubit is used further in the quantum computation. This continuation qubit $C$ needs to be in the state $HR^z(\alpha)\ket+$ which implies that a measurement of $T$ in $\ket{-\alpha}$ requires an additional $X$-gate operation. Hence, we need to reliably identify the state that qubit $C$ was projected into, due to the measurement on qubit $T$. Without projection errors, qubit $T$ would be in the state $\ket b \in\{\ket{\pm\alpha}\}$ after the measurement, and the $X$-gate on qubit $C$ needs to be inserted if and only if $\ket{b}=\ket{-\alpha}$. However, with projection errors, there are probabilities of projecting into the false states $\ket{\alpha}\otimes XHR^z(\alpha)\ket{+}$ which needs an application of $X$ even though qubit $T$ is in the $\ket{\alpha}$ state, and $\ket{-\alpha}\otimes HR^z(\alpha)\ket+$ which does not need an application of $X$ even though qubit $T$ is in the $\ket{-\alpha}$ state. Hence, we cannot identify the state of qubit $C$ by only looking at qubit $T$ after the measurement has taken place.

Instead, we introduce a verification register $V$, which is initialized in the state $\ket{0\ldots0}$. Before the measurement of target $T$, we have to initialize the graph state and the verification register $V$ in the state:
\begin{equation}
\label{eq:verf_scheme_state}
\begin{split}
    \ket{\tilde{\psi}}_{TCV} = &\ket\alpha\otimes HR^z(\alpha)\ket+\otimes\ket\alpha^{\otimes\#V}\\
    &+\ket{-\alpha}\otimes XHR^z(\alpha)\ket+\otimes\ket{-\alpha}^{\otimes\#V}.
\end{split}
\end{equation}
Therefore, we have to first rotate qubit $T$ back into the computational basis $\{\ket{0}, \ket{1}\}$. This is achieved by the unitary $HR^z(\alpha)$. Next, we will entangle qubit $T$ with the whole verification register $V$ by applying $\CNOT$s between qubit $T$ and all qubits in $V$. Finally, we rotate all qubits ($T$ and $V$) back to the measurement basis $\{\ket{\pm\alpha}\}$, this is achieved by applying $R^z(-\alpha)H$ to all qubits. The whole unitary to initialize $T+V$ in the desired state is thus given by
\begin{equation}
\label{eq:verf_scheme_unitary}
    (R^z(-\alpha)H)_T(R^z(-\alpha)H)_V\CNOT(T,V)(HR^z(\alpha))_T,
\end{equation}
from which we obtain the desired state in Eq.~\eqref{eq:verf_scheme_state}. Note that an essential part of the state preparation in Eq.~\eqref{eq:verf_scheme_unitary} is the generation of a GHZ state between qubit $T$ and the verification qubits in register $V$, $\CNOT(T,V)H_T$. This is a challenging task on superconducting qubit hardware~\cite{Baeumer2024, Liao2025}. For other physical platforms such as ion traps or neutral atoms, promising, very efficient protocols have been proposed by employing global $ZZ$-interaction gates~\cite{Yin2025, Zhang2026}. Using those protocols, the preparation of the entangled state between qubit $T$ and register $V$ in our mitigation method may therefore also be implemented with fewer operations and higher fidelities on such machines. This could suppress error propagation of repeated CNOT gates and reduce overall execution time.

After the projective measurement of qubit $T$, the verification register is then in the state $\ket\alpha^{\otimes\#V}$ or $\ket{-\alpha}^{\otimes\#V}$ and we need to apply an $X$ gate to qubit $C$ if and only if the verification state is $\ket{-\alpha}^{\otimes\#V}$. We proceed with measuring each verification qubit in the $\ket{\pm\alpha}$-basis, where again projection errors might happen. Note that if the measurements in the rotated basis are implemented via rotating the qubits into the computational basis and then performing computational basis measurements, we can simplify the unitary of Eq.~\eqref{eq:verf_scheme_unitary} and integrate it into the measurement process. In this case we simply need to apply $\CNOT(T,V)$ before the computational basis measurements of qubit $T$ and the register $V$. In order to determine the state of qubit $C$ after all measurements, we apply a majority voting protocol to the measurement outcomes of qubit $T$ and register $V$. The remaining misidentification probability for this protocol is thereby similar to Eq.~\eqref{eq:misident_readout_err} if we translate $N$ to the number of measured qubits, i.e., $N=\#V + 1$. The readout error is translated to the projection error or combined measurement error. We will use the notation $m_{\neg b,b}$ to denote the measurement error of the qubits in the following. This error can either be associated with a pure projection error or a combined measurement error.
Note that the proposed method can also be easily extended to multi-qubit measurements, if the measurements are implemented via unitary basis changes followed by single-qubit measurements. In this setup every single-qubit measurement can be mitigated as described above. An interesting future direction would, however, be the theoretical analysis of the misidentification rate of our method if we consider correlated measurement errors.

\section{Considering CNOT errors in the verification scheme}
\label{sect:cnot_errs}
For simplicity, we will consider pure-state preserving errors. If the \CNOT-gate were error-free, then $\CNOT \ket{+}\otimes\ket{0}=\frac{1}{\sqrt{2}}(\ket{0}\otimes\ket{0}+ \ket{1}\otimes\ket{1})=\ket{\Phi^+}$ and $\CNOT \ket{-}\otimes\ket{0}=\frac{1}{\sqrt{2}}(\ket{0}\otimes\ket{0} - \ket{1}\otimes\ket{1})=\ket{\Phi^-}$ should hold. However, if the $\CNOT$ has an error rate $\gamma_{\Phi^\mp,\Phi^\pm}$, then the final state of the verification qubit is 
\begin{equation*}
  \sqrt{1-\gamma_{\Phi^-,\Phi^+}}\ket\Phi^++\sqrt{\gamma_{\Phi^-,\Phi^+}}\ket{\Phi^-}
\end{equation*}
or
\begin{equation*}
  \sqrt{1-\gamma_{\Phi^+,\Phi^-}}\ket{\Phi^-}+\sqrt{\gamma_{\Phi^+,\Phi^-}}\ket{\Phi^+}.
\end{equation*}
Either way, there is a probability $\gamma$ that the qubits are not correctly entangled. For the entanglement between the verification register and the OWQC qubit, we thus obtain an accumulated error rate $\tilde{\gamma}$ that a verification qubit is in the wrong state before its measurement. If we add the measurement error $m_{\neg b, b}$ for projection (and readout), then there is a probability of $(1-m_{\neg b, b})(1-\tilde{\gamma})+m_{\neg b, b}\tilde{\gamma}=1-m_{\neg b, b}-\tilde{\gamma}+2m_{\neg b, b}\tilde{\gamma}$ that the state is verified correctly and a probability of $(1-m_{\neg b, b})\tilde{\gamma}+m_{\neg b, b}(1-\tilde{\gamma})=m_{\neg b, b}+\tilde{\gamma}-2m_{\neg b, b}\tilde{\gamma}$ that the state is incorrectly identified.

\begin{figure}
    \centering
    \begin{subfigure}{0.45\textwidth}
        \centering
        \includegraphics[width=\textwidth]{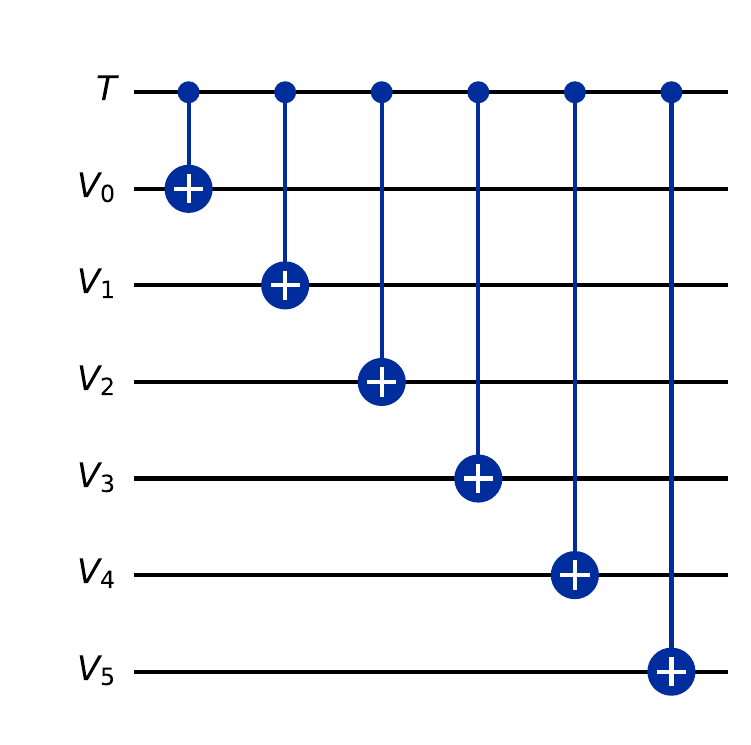} 
        \captionsetup{justification=centering}
        \caption{}
        \label{fig:circuit_linear_log_a}
    \end{subfigure}
    \hfill
    \begin{subfigure}{0.45\textwidth}
        \centering
        \includegraphics[width=\textwidth]{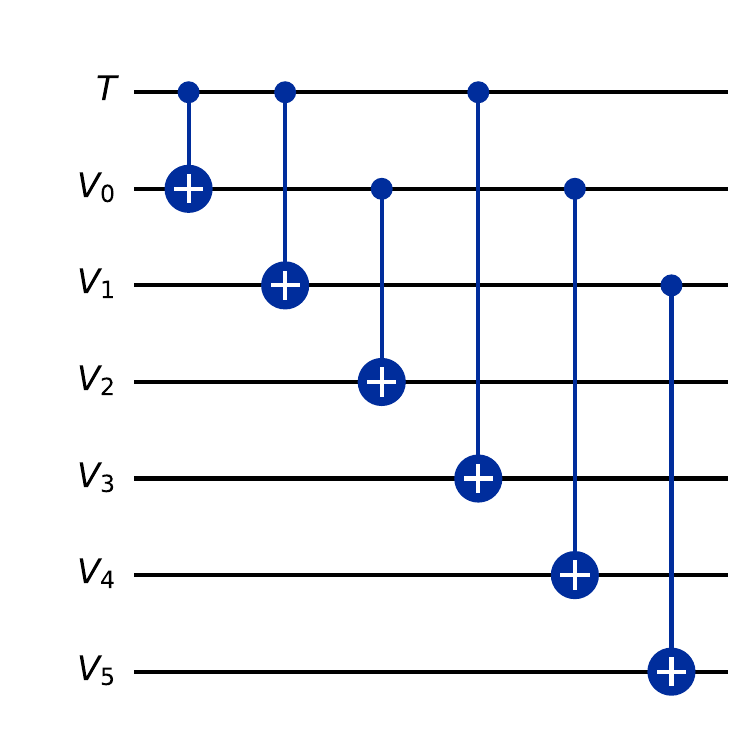} 
        \captionsetup{justification=centering}
        \caption{}
        \label{fig:circuit_linear_log_b}
    \end{subfigure}
    \caption{Circuit diagram examples of the linear (panel (a)) and the log-depth (panel (b)) $\CNOT$ entangling protocol between a OWQC target qubit $T$ and a verification register $V$.}
    \label{fig:circuit_linear_log}
\end{figure}

For a linear $\CNOT$ setup, where a $\CNOT$ gate is applied between each verification qubit and the target qubit (cf. Fig.~\ref{fig:circuit_linear_log_a}), the first-order approximation for the accumulated error in the small-$\gamma$ regime is $\tilde{\gamma}\approx(N-1)\gamma$ because the $\CNOT$ error can also influence the target qubit $T$ in each application. In this case, we are again in the same protocol as the projection-free readout, but we have the larger effective error rate $m_{\neg b, b}+(N-1)\gamma-2m_{\neg b, b}(N-1)\gamma$.

If we use a log-depth verification setup (cf. Fig.~\ref{fig:circuit_linear_log_b}), that is, populating the verification register according to the following $S=\log N$ steps
\begin{samepage}
\begin{enumerate}
\item[1:] $\CNOT(T,V_1)$
\item[2:] $\CNOT(T,V_2)\CNOT(V_1,V_3)$
\item[] $\vdots$
\item[$S$:] $\CNOT(T,V_{2^{S-1}})\prod_{j=1}^{2^{S-1}-1}\CNOT(V_j,V_{2^{S-1}+j})$
\end{enumerate}
\end{samepage}
then we have accumulating effects of the error $\gamma$ to a maximum depth of $\log N$. Fig.~\ref{fig:circuit_linear_log} shows a circuit diagram example for the linear and the log-depth protocol. 

If we model our erroneous $\CNOT$ via the action of
\begin{align*}
  \CNOTp=
  \begin{pmatrix}
    1-\gamma_{\Phi^-,\Phi^+}&\gamma_{\Phi^+,\Phi^-}\\
    \gamma_{\Phi^-,\Phi^+}&1-\gamma_{\Phi^+,\Phi^-}\\
  \end{pmatrix}
  =1+\Gamma
\end{align*}
in propagating the error, then we have depth errors of 
\begin{align*}
  \CNOTp^k=&(1+\Gamma)^k=\sum_{j=0}^k{\binom{k}{j}}\Gamma^j.
\end{align*}

\begin{remark}
    It should be noted that the $\CNOT$ error modelled via $\Gamma$ here is more general than the pure-state error channel used above to obtain the form of $\Gamma$. $\Gamma$ describes an effective transition matrix where each $\ket{\Phi^\pm}$ state remains in $\ket{\Phi^\pm}$ with probability $1-\gamma_{\Phi^\mp,\Phi^\pm}$ and maps to the other state with probability $\gamma_{\Phi^\mp,\Phi^\pm}$. Importantly, non-coherent error channels can be incorporated into $\Gamma$ as well. For example, a depolarizing channel of strength $\lambda$ can be considered as a probability $\lambda$ for the states $\ket{\Phi^\pm}$ to be mapped to an equal probability of either outcome, i.e., $\Gamma=\frac{\lambda}{2}\begin{pmatrix}
        -1&1\\1&-1
    \end{pmatrix}$. 
\end{remark}

The error propagation depends on the operator norm (spectral norm) $\lVert\Gamma\rVert=\gamma_{\Phi^-,\Phi^+}+\gamma_{\Phi^+,\Phi^-}$ of $\Gamma$. If $\lVert\Gamma\rVert$ is reasonably small, say in the $1\%$ range, then $\lVert\Gamma^2\rVert$ is in the $0.1$\textperthousand{} range, and so realistically, the first-order propagated error to depth $k$ is approximately
\begin{align*}
  \CNOTp^{k}\approx1+k\ \Gamma.
\end{align*}
and
\begin{equation*}
    \tilde{\gamma}\approx k\gamma.
\end{equation*}

For the linear protocol we get a maximum accumulation depth of $k=N-1$ and for the log-depth protocol we get $k=\log_2 N$.
In other words, a starting error of about $1\%$ with a size 64 register leads to an effective maximum error of about $6\%$ for the log-depth protocol. The remaining misidentification probability is then $\eps \approx 2.53\cdot 10^{-23}$ according to Eq.~\eqref{eq:misident_readout_err}.
Thus, choosing a log-depth verification setup with register size $\# V$ and $N=\# V +1$, the effective error becomes 

\begin{equation}
\label{eq:eff_error}
    \tilde{m}_{\neg b, b} = m_{\neg b, b}+(1-2m_{\neg b,b})(\log_2 N)\gamma,
\end{equation}

and the probability of misidentification becomes

\begin{equation}
\label{eq:misident_logV}
  \eps=I_{\tilde{m}_{\neg b, b}}(N-\lfloor N/2\rfloor,1+\lfloor N/2\rfloor).
\end{equation}

Note that the register size and the effective error rate are now coupled according to Eq.~\eqref{eq:eff_error}. This means that for a given initial error rate $m_{\neg b,b}$ and a given $\CNOT$ error $\gamma$, the effective error depends on the register size. In order to thus plot $\eps(\tilde{m}_{\neg b, b})$, $N$ will not necessarily be an integer and we will use an upper bound estimate for $I_{\tilde{m}_{\neg b, b}}(N-\lfloor N/2\rfloor,1+\lfloor N/2\rfloor)$ which avoids any floor function and allows for non-integer $N$. This estimate is given by

\begin{equation}
\label{eq:misident_logV_est}
\begin{split}
    \eps_{\text{est}}=&I_{\tilde{m}_{\neg b, b}}(N/2,1+N/2)\\
    \geq& I_{\tilde{m}_{\neg b, b}}(N-\lfloor N/2\rfloor,1+\lfloor N/2\rfloor)=\eps.
    \end{split}
\end{equation}
Additionally, $\CNOT$ errors acting on one qubit can propagate to all its entanglement partners and thus lead to correlated effective errors. This means that modeling the misidentification probability based on a binomial outcome distribution of the measurements (cf. Sec.~\ref{sect:compensating_readout_errs} and Eq.~\eqref{eq:misident_logV}) is no longer accurate. The resulting regularized incomplete beta function $I_{\tilde{m}_{\neg b, b}}(N-\lfloor N/2\rfloor,1+\lfloor N/2\rfloor)$ thus only approximates the misidentification probability $\eps$ if the correlations between the effective measurement errors remain small.

\begin{figure}
    \centering
    \includegraphics[width=0.9\textwidth]{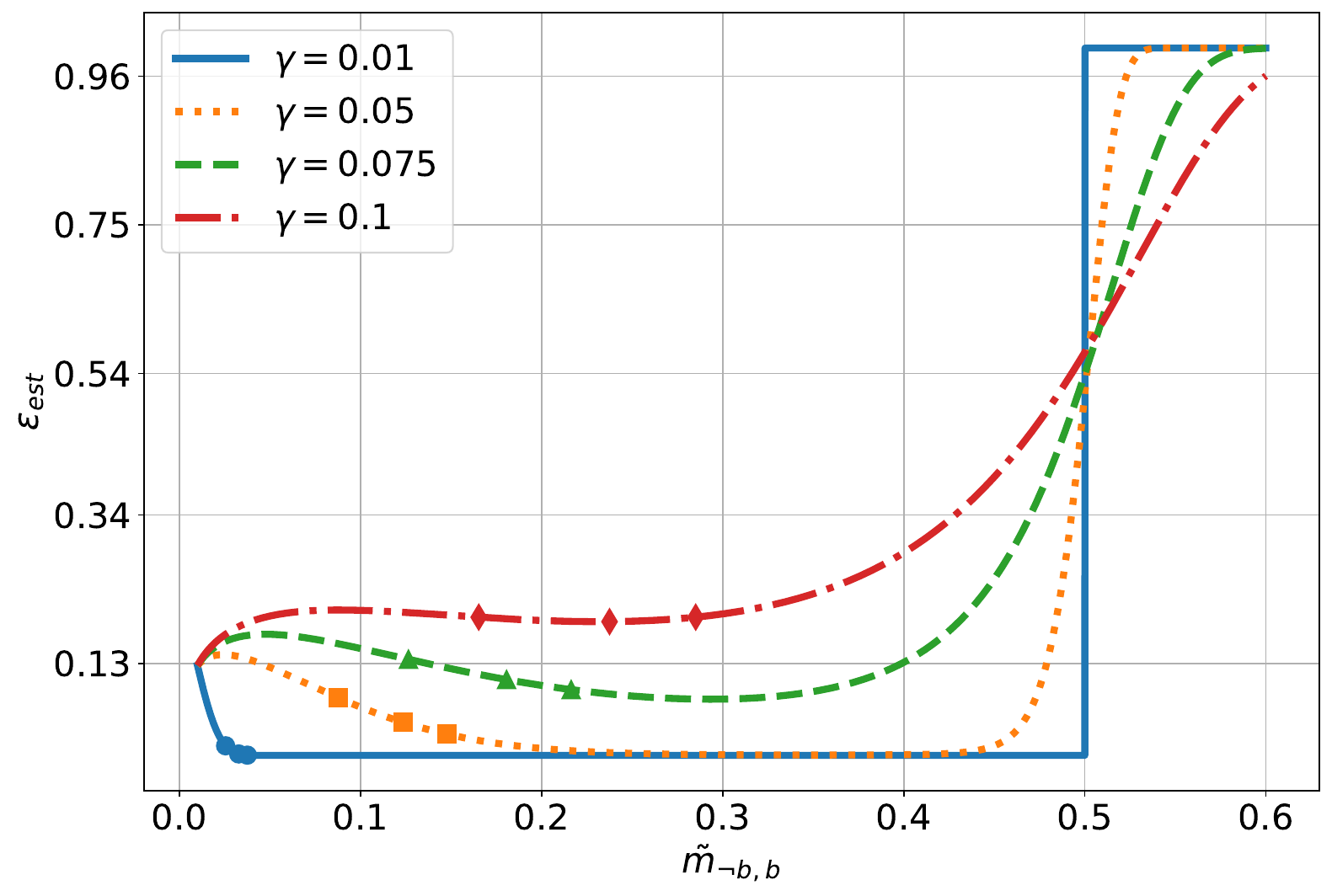}
    \caption{Estimate of the misidentification probability $\eps_{\text{est}}$ from Eq.~\eqref{eq:misident_logV_est} in dependence of an effective error $\tilde{m}_{\neg b, b} = m_{\neg b, b}+(1-2m_{\neg b, b})(\log_2 N)\gamma$ for different $\CNOT$ error rates $\gamma$. The initial error rate is set to $m_{\neg b,b}=0.01$. The blue circle, orange square, green triangle and red diamond markers indicate the resulting $\eps_{\text{est}}$ for $N=[3,5,7]$ and the corresponding $\gamma$.}
    \label{fig:misident_by_effective_error}
\end{figure}

\begin{figure}
    \centering
    \includegraphics[width=0.9\textwidth]{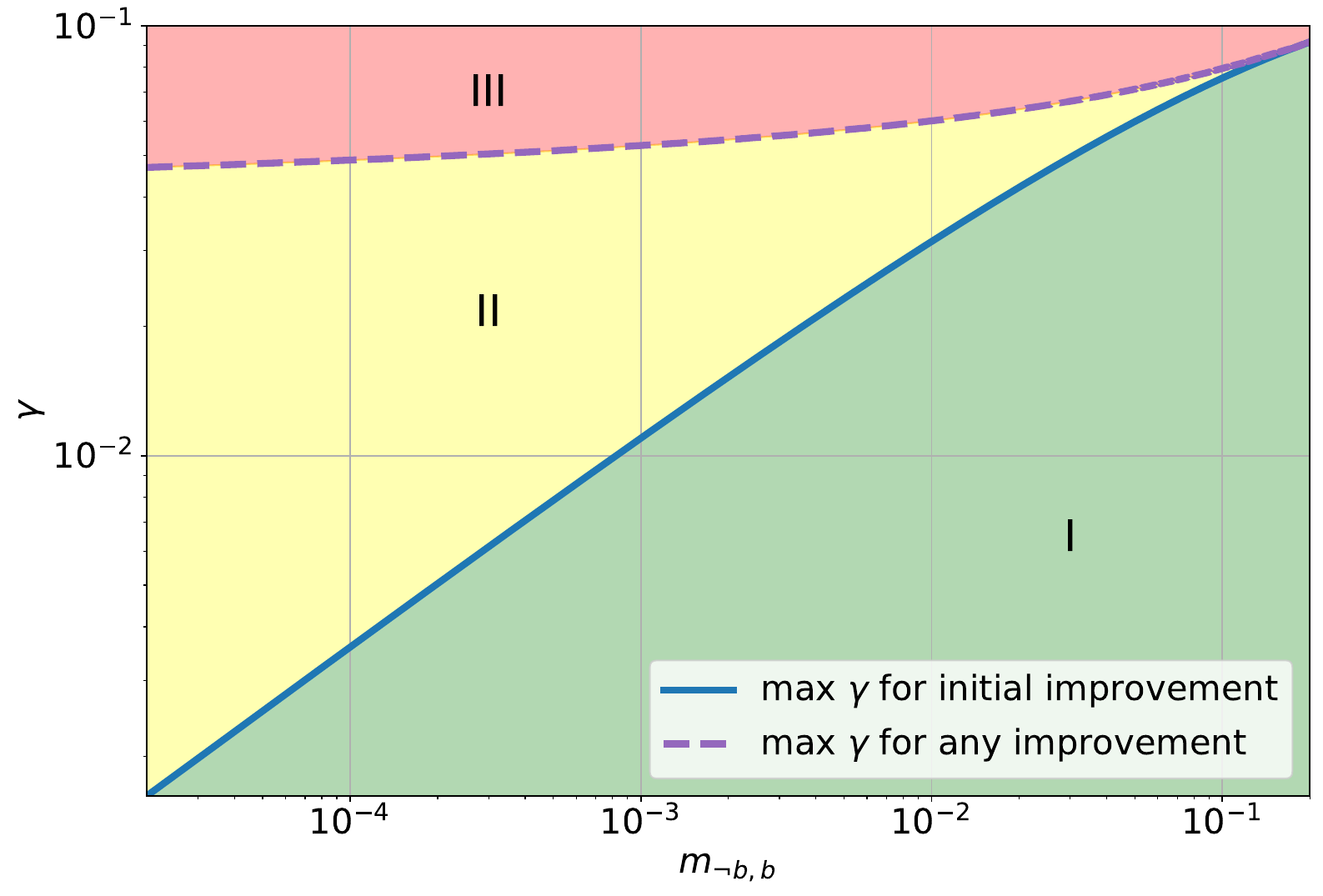}
    \caption{Critical $\CNOT$ errors $\gamma$ which separate the regimes of initial improvement (III) (cf. Eq.~\eqref{eq:regimes_initial_improvement}), initial worsening but then improvement (II) (cf. Eq.~\eqref{eq:regimes_any_improvement}) and no improvement (I) in the misidentification probability $\eps$, shown in Fig.~\ref{fig:misident_by_effective_error}, in dependence of the initial error rate $m_{\neg b, b}$. The solid blue line shows the critical $\gamma$ which separates the initial improvement (III) from the initial worsening but then improvement regime (II). The dashed purple line shows the critical $\gamma$ which separates the initial worsening but then improvement regime (II) from the no improvement regime (I).
    }
    \label{fig:regimes}
\end{figure}

Fig.~\ref{fig:misident_by_effective_error} shows the estimate of the misidentification probability $\eps_{\text{est}}$ in dependence of the effective error rate $\tilde{m}_{\neg b, b}$ from Eq.~\eqref{eq:misident_logV_est}. The function is plotted for different $\CNOT$ errors $\gamma$. Even for small values, e.g., $\gamma = 0.01$, we see that the misidentification probability can jump to $1.0$ if the effective error reaches $50\%$. The protocol thus fails for effective errors $\tilde{m}_{\neg b, b}\geq50\%$, i.e., for too large register sizes. For small effective errors, a small $\CNOT$ error $\gamma$ yields an immediate improvement (cf. (III) in Fig.~\ref{fig:regimes}). Midsize $\CNOT$ errors can result in an initial worsening before an improvement is achieved for larger register sizes (cf. (II) in Fig.~\ref{fig:regimes}). Large $\CNOT$ errors will yield no improvement at all (cf. (I) in Fig.~\ref{fig:regimes}).
By iteratively calculating $\eps_{N=3} - \eps_{N=1}$ and $\eps_{N\in[1, 1001]} - \eps_{N=1}$ for different $\gamma$ values and a fixed initial error rate $m_{\neg b, b}$ (cf. Eq.~\eqref{eq:misident_logV}), we can extract the critical $\gamma$ values that separate the three regimes as shown in Fig.~\ref{fig:regimes}. Initial improvement is thereby defined as
\begin{equation}
    \label{eq:regimes_initial_improvement}
    \eps_{N=3} - \eps_{N=1} < 0 \qquad \text{(initial improvement)},
\end{equation}
and any improvement is defined as
\begin{equation}
    \label{eq:regimes_any_improvement}
    \eps_{N\in[1, 1001]} - \eps_{N=1} < 0 \qquad \text{(any improvement)}.
\end{equation}

Note that additional single-qubit gate errors during the preparation of the state $\ket{\tilde{\psi}}_{TCV}$ in Eq.~\eqref{eq:verf_scheme_state} from the initial graph state can also be incorporated into the error rate $\gamma$.

\section{Simulation and Hardware Results}
\label{sect:simulation_res}
\begin{figure}
    \centering
    \includegraphics[width=0.9\textwidth]{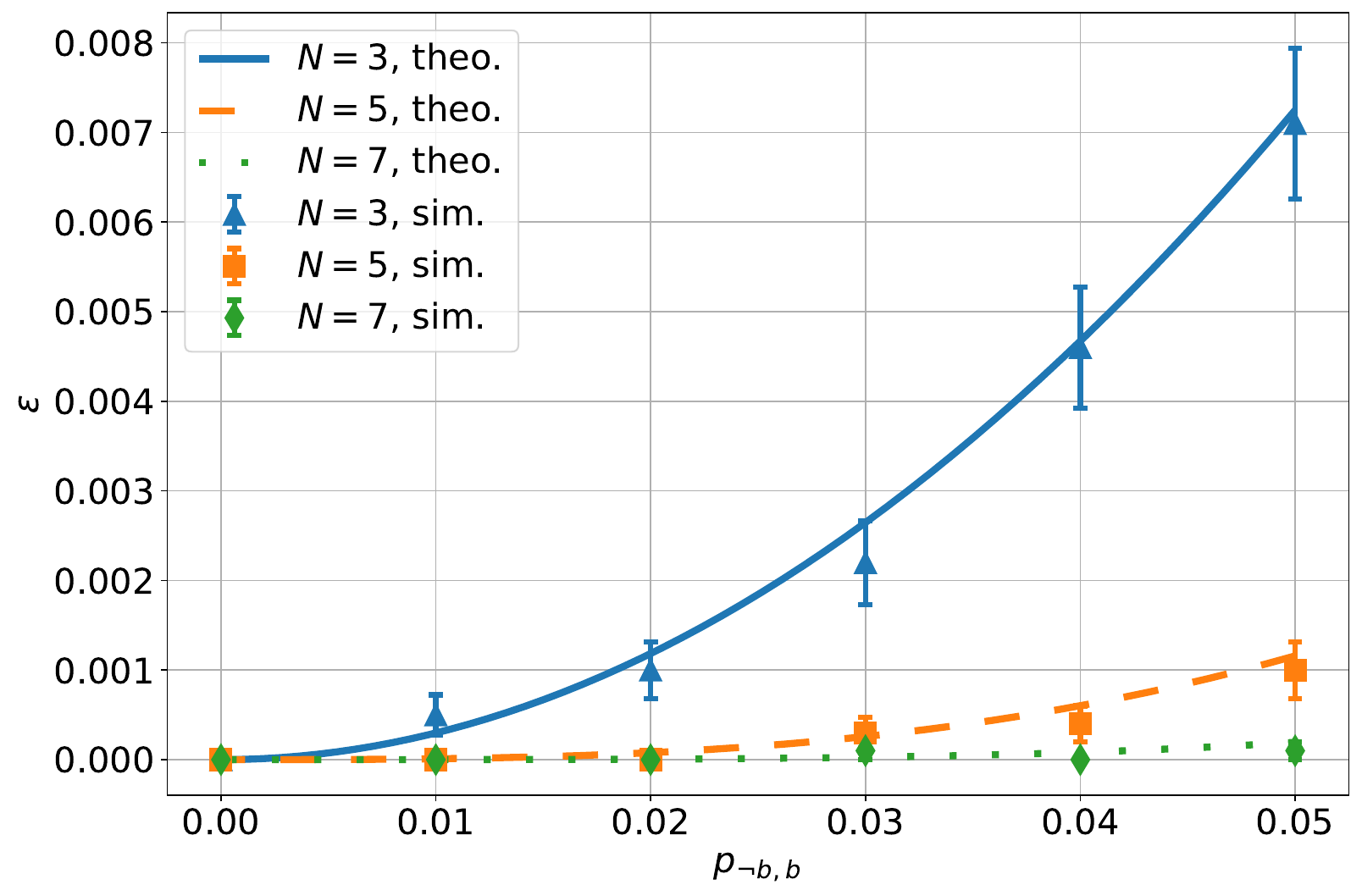}
    \caption{Mean misidentification rate $\eps$ of the OWQC shown in Eq.~\eqref{eq:verf_scheme_state} in dependence of the projection error probability $p_{\neg b, b}$ of the single-qubit measurement. The OWQC was executed $10000$ times for each error rate. The misidentification rate was calculated for $10$ different batches, each containing $1000$ runs, and afterwards averaged over those $10$ batches. The different markers show the results for different sizes of the verification register, together with their standard error of the mean for the batch average. The different lines show the corresponding theoretical predictions using Eq.~\eqref{eq:misident_readout_err}. A misidentification corresponds to a resulting state after measurement with less than $50\%$ overlap with the desired state.}
    \label{fig:owqc_simulation_results}
\end{figure}

\begin{figure}
    \centering
    \begin{subfigure}{0.45\textwidth}
        \centering
        \includegraphics[width=\linewidth]{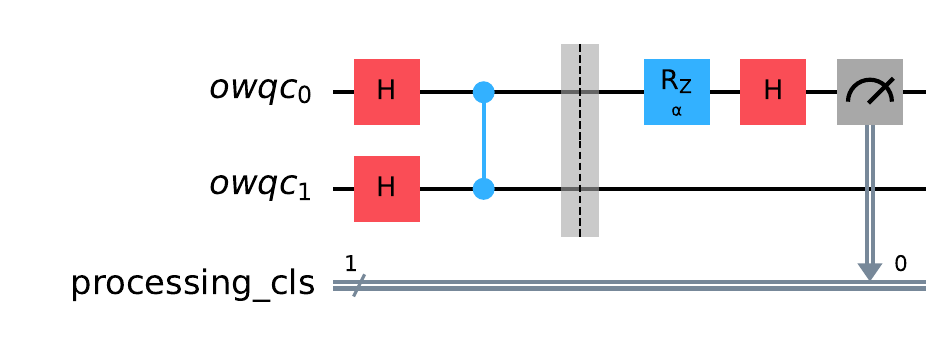} 
        \captionsetup{justification=centering}
        \caption{}
        \label{fig:owqc_circuits_qiskit_a}
    \end{subfigure}
    \begin{subfigure}{0.45\textwidth}
        \centering
        \includegraphics[width=\linewidth]{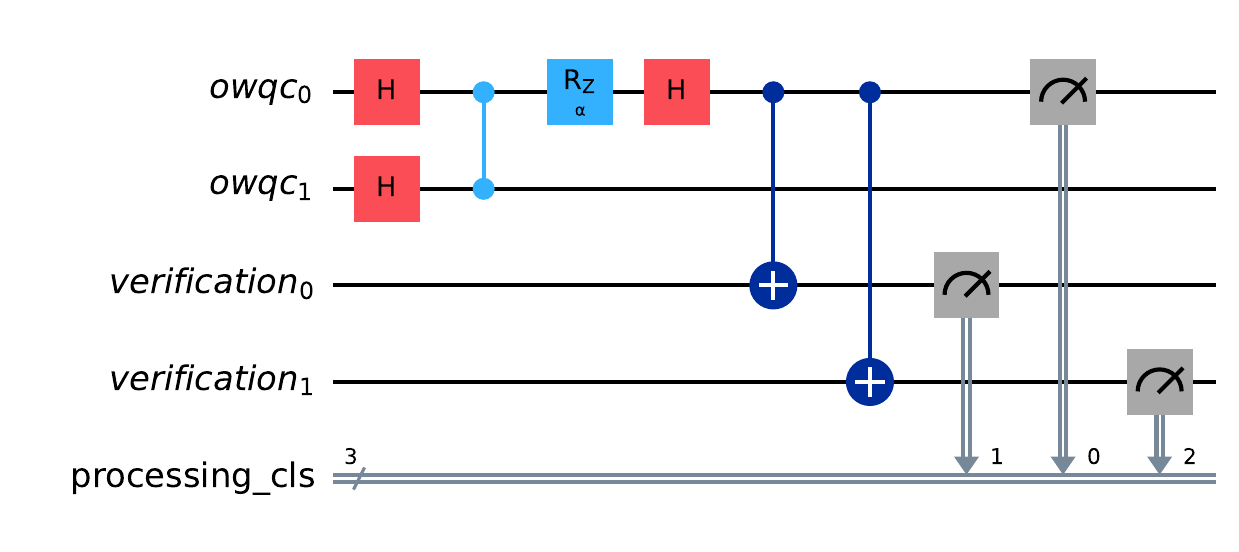} 
        \captionsetup{justification=centering}
        \caption{}
        \label{fig:owqc_circuits_qiskit_b}
    \end{subfigure}
    \begin{subfigure}{0.9\textwidth}
        \centering
        \includegraphics[width=\linewidth]{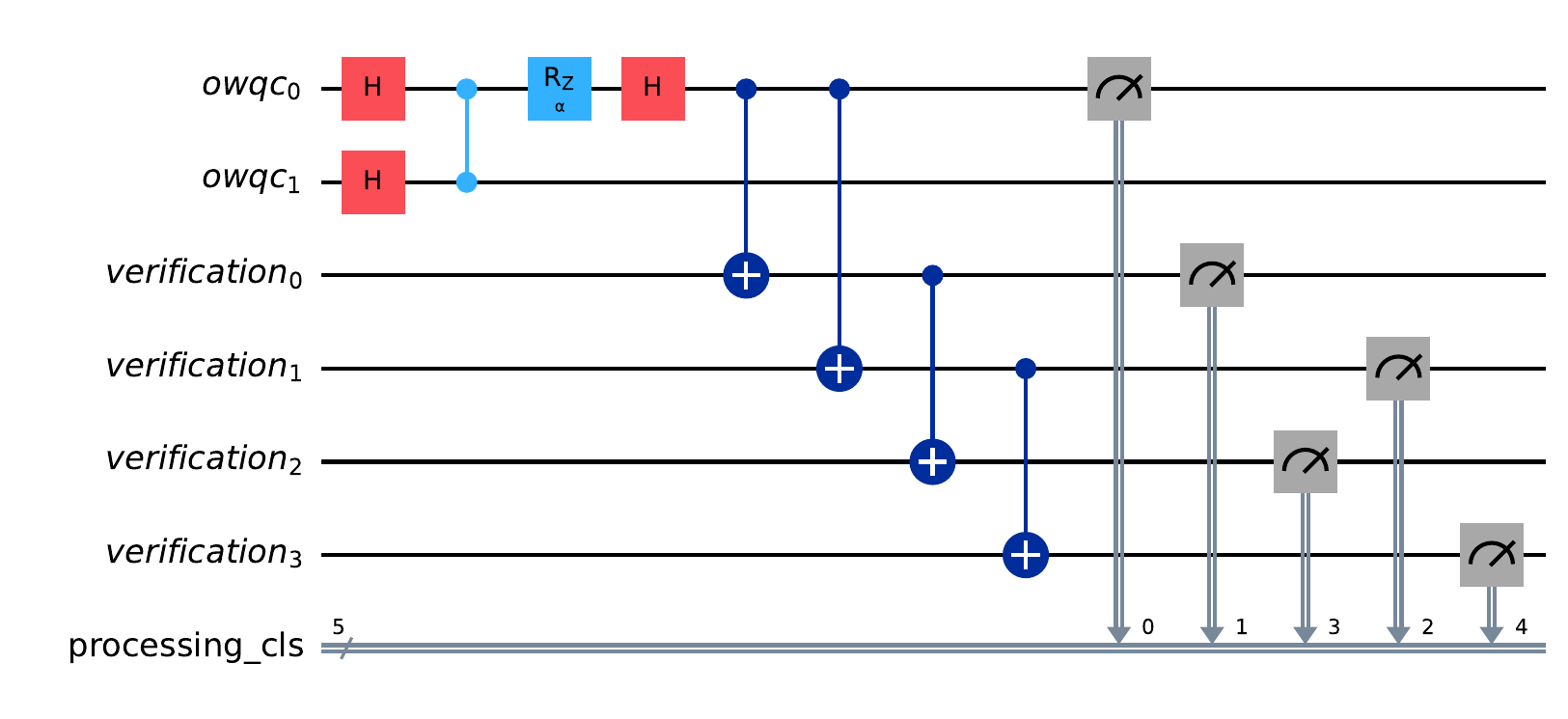} 
        \captionsetup{justification=centering}
        \caption{}
        \label{fig:owqc_circuits_qiskit_c}
    \end{subfigure}
    \caption{Qiskit circuits executed on the IBM quantum hardware \textsc{ibm\_brussels}. Panel (a): The upper circuit contains the circuit representation of the graph state generation shown in Eq.~\eqref{eq:HRz_owqc} followed by a measurement on the qubit $owqc_0$ in the rotated basis $\{\ket{\pm\alpha}\}$. The two parts are separated by a barrier in the circuit. Panel (b): The middle circuit is the extension of the upper circuit by using additionally the proposed mitigation scheme with two verification qubits. Panel (c): The lower circuit is the extension of the upper circuit by using additionally the proposed mitigation scheme with four verification qubits and including the error propagation as it appears in the log-depth implementation.}
    \label{fig:owqc_circuits_qiskit}
\end{figure}

\begin{figure}
    \centering
    \includegraphics[width=0.9\textwidth]{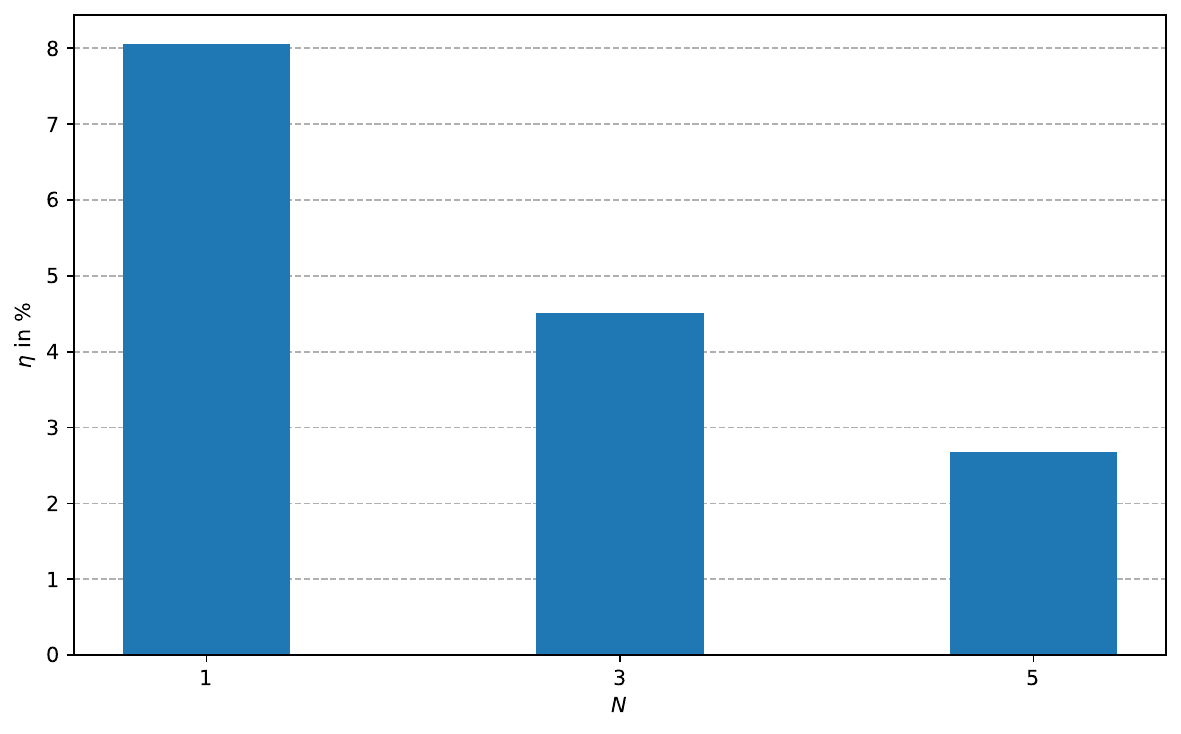}
    \caption{Relative error $\eta$ (cf. Eq.~\eqref{eq:rel_error}) of the circuits shown in Fig.~\ref{fig:owqc_circuits_qiskit}, which have been executed on \textsc{ibm\_brussels}. The obtained outcome distributions have been averaged over $8$ different circuit instances of Fig.~\ref{fig:owqc_circuits_qiskit}, corresponding to $8$ randomly chosen rotation angles $\alpha\in [0,2\pi)$. Each circuit was executed with $10000$ shots. The total number of measured qubits $N$ includes the to-be-measured qubit of the initial computation and all verification qubits. The unmitigated measurement error of the to-be-measured qubit (baseline error) is thus shown as $N=1$ in the plot.}
    \label{fig:ibm_hardware_run}
\end{figure}

In this section, we want to illustrate how the proposed verification scheme behaves for the elementary compute step of a OWQC, implementing the gate $HR^z(\alpha)$ acting on $\ket{+}$ (cf. Sec.~\ref{sect:proj_error_mitigation}). For this, we first classically simulate the OWQC with our verification scheme, considering only projection errors as discussed above and no readout errors. This means we assume that projection and readout can be separated in this case. The verification scheme to compensate the projection errors leads to a similar voting protocol as for the readout errors with the same predicted probability of misidentification (cf. Eq.~\eqref{eq:misident_readout_err}).
Overall, the simulations are in very good agreement with the theoretical predictions as shown in Fig.~\ref{fig:owqc_simulation_results}. We tested the verification scheme for different register sizes, ranging from $N=3$ to $N=7$ qubits and different projection error probabilities, ranging from $0\%$ to $5\%$ error for any qubit projection.
Additionally, we also tested our mitigation scheme on the IBM quantum hardware \textsc{ibm\_brussels}. In this case, projection and readout cannot be separated and we therefore have to mitigate the combined error with our verification scheme. We implemented the generation of the state from Eq.~\eqref{eq:verf_scheme_state} followed by the measurements of the OWQC qubit and the verification register $V$ in a quantum circuit using Qiskit (cf. Fig.~\ref{fig:owqc_circuits_qiskit}). Note that this implementation represents the elementary compute step in every OWQC. By applying the verification scheme at every step of the ongoing OWQC, we can mitigate measurement errors in real-time.
In the ideal case, without any errors, both outcomes for the measured OWQC qubit ($owqc_0$ in Fig.~\ref{fig:owqc_circuits_qiskit}) are equally probable, $\mathrm{Pr}(0)=\mathrm{Pr}(1)=0.5$, independent of the rotation angle $\alpha$. Measurement errors will shift this outcome distribution towards a higher probability of measuring $0$ or measuring $1$. Our mitigation scheme should thus shift the erroneous outcome distribution back towards the ideal distribution. We calculated the relative error $\eta$ of the measured (erroneous) outcome distribution $\mathrm{Pr}_{meas}(0)$, $\mathrm{Pr}_{meas}(1)$ compared to the ideal outcome,
\begin{equation}
\label{eq:rel_error}
    \eta = \frac{\lvert\mathrm{Pr}_{meas}(0)-0.5\lvert}{0.5}=\frac{\lvert\mathrm{Pr}_{meas}(1)-0.5\lvert}{0.5}.
\end{equation}
We expect that this error will decrease for increasing verification register size. This expectation is confirmed by the obtained hardware results as presented in Fig.~\ref{fig:ibm_hardware_run}. We observe that the relative error is almost halved for each increase in the verification register size.

\section{Conclusion}
\label{sect:conclusion}
We first introduced a measurement error model that, in contrast to most previous works, makes a clear distinction between errors that occur through the readout apparatus after projecting the qubit during the measurement (readout error) and the actual projection error that occurs during the projection. Note that for cases where those two error sources are not clearly separable, we can treat the combined measurement error similarly to a pure projection error in our model and still apply our mitigation method. We then discussed the compensation of the readout error through a voting protocol in great detail and gave theoretical predictions on the number of consecutive readouts for given error rates and its resulting probability to misidentify the post-measurement state, which vanishes quickly with small effort in terms of the readout number. Next, we discussed the compensation of projection errors or inseparable, combined measurement errors (readout + projection), by using a register of verification qubits that are entangled with the next to-be-measured qubit in the OWQC. Measuring the to-be-measured qubit as well as all verification qubits then allows us to apply a similar voting protocol as for the readout errors on the measurement outcomes. The derived theoretical predictions are the same, simply replacing the readout error with the projection error or combined measurement error and the number of readouts with the number of measured qubits. Note that, by re-using the verification qubits after each measurement, we require only a constant number of verification qubits depending on the measurement error rates. Finally, we showed in proof-of-principle simulations and experiments executed on real IBM quantum hardware \textsc{ibm\_brussels}, that our method works very well and can be used to mitigate measurement errors in real-time during a OWQC.

A future next step could be the extension of these proof-of-principle experiments to the implementation of more complex (multi-qubit) logical gates and small algorithms as OWQCs and applying our mitigation protocol in those computations. Here, a benchmark among different random logical unitaries might be valuable to assess the general performance.

Further, our analysis assumes that the measurement errors of the target qubit $T$ and the verification qubits in $V$ are statistically independent, which is what lets the misidentification probability of Eq.~\eqref{eq:misident_readout_err} decrease so rapidly with $N$. In realistic implementations this holds only approximately, due to correlated measurement errors between the entangled qubits. Under such conditions the outcome distribution is no longer a binomial distribution (cf. Eq.~\eqref{eq:binomial_readout_dist}) and Eq.~\eqref{eq:misident_readout_err} is no longer valid. A quantitative analysis of the misidentification probability with correlated measurement errors is left for future work.
\backmatter

\bmhead{Acknowledgements}

We thank Ferdinand Schmidt-Kaler and his group for helpful discussions and the provided hardware data of measurement error rates. This work is funded by the European Union’s Horizon Europe Framework Programme (HORIZON) under the ERA Chair scheme with grant agreement no.\ 101087126 and by the Deutsche Forschungsgemeinschaft (DFG, German Research Foundation) – Project-ID 429529648 – TRR 306 QuCoLiMa (“Quantum Cooperativity of Light and Matter’’).
This work is supported with funds from the Ministry of Science, Research and Culture of the State of Brandenburg within the Centre for Quantum Technologies and Applications (CQTA). 
\begin{center}
    \includegraphics[width = 0.08\textwidth]{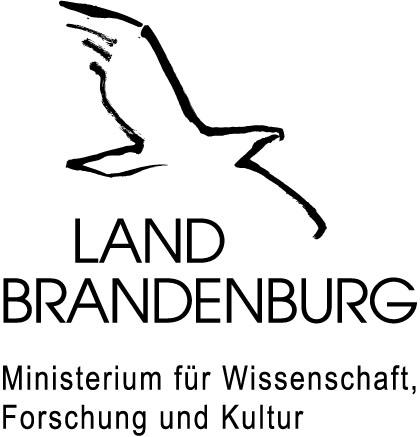}
\end{center}
\FloatBarrier

\section*{Declarations}
\subsection{Funding}
This work is funded by the European Union’s Horizon Europe Framework Programme (HORIZON) under the ERA Chair scheme with grant agreement no.\ 101087126 and by the Deutsche Forschungsgemeinschaft (DFG, German Research Foundation) – Project-ID 429529648 – TRR 306 QuCoLiMa (“Quantum Cooperativity of Light and Matter’’).
This work is supported with funds from the Ministry of Science, Research and Culture of the State of Brandenburg within the Centre for Quantum Technologies and Applications (CQTA).

\subsection{Data availability}
The datasets of the numerical simulations, presented in Fig.~\ref{fig:owqc_simulation_results}, and the hardware results, presented in Fig.~\ref{fig:ibm_hardware_run}, are available from \url{https://dx.doi.org/10.22000/1ffk7e31t4dfmd9y}. 

\subsection{Code availability}
The code used to generate the numerical simulations (cf. Fig.~\ref{fig:owqc_simulation_results}) and execute the circuits of Fig.~\ref{fig:owqc_circuits_qiskit} on IBM hardware (cf. Fig.\ref{fig:ibm_hardware_run}) will be made available by the authors upon reasonable request.

\subsection{Author contribution}
T.H. conceived the original idea, developed the mitigation scheme and corresponding error model, and derived the resulting misidentification probability distributions. S.S. derived the formula to determine N in the polynomial approximation, and implemented the numerical simulations as well as the quantum hardware experiments via the Qiskit interface. T.H. and S.S. jointly wrote the manuscript and prepared the figures. K.J. and J.Z. provided technical guidance throughout the project and critically reviewed the manuscript.

\bibliography{papers}
\newpage
\begin{appendices}
\section{Derivation of \texorpdfstring{$N(\varepsilon)$}{TEXT} in the polynomial approximation}
\label{sect:N_eps_derivation}
The dependence of $N$ on $\varepsilon$ is shown in Eq.~\eqref{eq:N_eps} in the main text. In the following, a detailed derivation of this equation is presented.

The regularized incomplete beta function from Eq.~\eqref{eq:misident_readout_err}, modelling the misidentification probability, can be written in the polynomial approximation shown in Eq.~\eqref{eq:beta_poly_approx} as:
\begin{equation}
\label{eq:misident_readout_err_poly_approx}
\begin{aligned}
    \eps &= I_{r_{\neg b, b}}\left(N-\left\lfloor \frac{N}{2}\right\rfloor, 1+\left\lfloor \frac{N}{2}\right\rfloor\right) \\ 
    &\approx \frac{\Gamma(N+1)}{(N-\left\lfloor \frac{N}{2}\right\rfloor)\Gamma(N-\left\lfloor \frac{N}{2}\right\rfloor)\Gamma(1+\left\lfloor \frac{N}{2}\right\rfloor)}r_{\neg b, b}^{N-\left\lfloor \frac{N}{2}\right\rfloor}\\
    &=\frac{N!}{(N-\left\lfloor \frac{N}{2}\right\rfloor)!(\left\lfloor \frac{N}{2}\right\rfloor)!}r_{\neg b, b}^{N-\left\lfloor \frac{N}{2}\right\rfloor}\\
    &= {\binom{N}{\left\lfloor \frac{N}{2} \right\rfloor}}r_{\neg b, b}^{N-\left\lfloor \frac{N}{2}\right\rfloor}
\end{aligned}
\end{equation}
Assuming that $N$ is an odd number, since the voting protocol would be indefinite otherwise, we can write the floor function $\lfloor N/2 \rfloor$ as:
\begin{equation*}
    \left\lfloor \frac{N}{2} \right\rfloor \underset{N \text{ is odd}}{=} \frac{N-1}{2}
\end{equation*}
The binomial coefficient in Eq.~\eqref{eq:misident_readout_err_poly_approx} can thus be simplified as follows, using Stirling's approximation for factorials $n! \approx \sqrt{2\pi n}(n/e)^n$:
\begin{align*}
    \binom{N}{\frac{(N-1)}{2}} &= \frac{N!}{\frac{N-1}{2}!\frac{N+1}{2}!}=\frac{N!}{\left(\frac{N+1}{2}!\right)^2\frac{2}{N+1}}\\
    &=\frac{N!(N+1)}{\left(\frac{N+1}{2}!\right)^2\cdot 2}\approx\frac{4^{\frac{(N+1)}{2}}}{\sqrt{2\pi (N+1)}}
\end{align*}
This yields a compact approximation for the misidentification probability:
\begin{equation}
\label{eq:misident_readout_err_striling_approx}
    \eps \approx \frac{(4r_{\neg b, b})^{\frac{N+1}{2}}}{\sqrt{2\pi(N+1)}}
\end{equation}
In order to invert this function $\eps(N)$ for a given fixed $r_{\neg b, b}$ and thus find $N(\eps)$ we can use the Lambert $W$-function. This function is defined as the inverse of $ye^y$,
\begin{equation*}
    ye^y=z \, \Leftrightarrow \, W(z) = y \quad \forall y,z\in\mathbb{R},\,z\geq-\frac{1}{e}.
\end{equation*}
First, we will rewrite Eq.~\eqref{eq:misident_readout_err_striling_approx} in the form
\begin{equation*}
    (2\pi)^{-1}\eps^{-2}=(N+1)R^{-(N+1)},
\end{equation*}
with $R=4r_{\neg b, b}$. Next, we apply $R^x=e^{x\ln (R)}$ and find the required form to use the Lambert $W$-function
\begin{align}
    -\frac{\ln(R)}{2\pi\eps^{2}} &= -(N+1)\ln(R)e^{-(N+1)\ln(R)} \nonumber\\
    \Rightarrow N(\eps) &= \frac{-W(-\frac{\ln(R)}{2\pi\eps^2})}{\ln(R)}- 1
\end{align}
Note that the argument of the Lambert $W(z)$ has to fulfill $z\geq -1/e$. Since $z=-1/e$ is the global minimum of $z=ye^y$. This is fulfilled for any $\eps$ if $r_{\neg b, b}\leq 0.25$,
\begin{equation*}
    \ln(R) = \ln(4r_{\neg b,b})\leq 0 \, \Leftrightarrow \, r_{\neg b, b}\leq 0.25.
\end{equation*}
For higher measurement errors, the polynomial approximation of the regularized incomplete beta function deviates significantly from the exact function. In this case, other methods to extract the required $N$ for a given $\eps$ can be employed, e.g., iteratively calculating $\eps$ from $I_{r_{\neg b, b}}(N-\lfloor N/2\rfloor,1+\lfloor N/2\rfloor)$ while increasing $N$.

\section{Detailed discussion of the \texorpdfstring{$\eps_{\mathrm{est}}$}{TEXT} regimes}
\begin{figure*}
    \centering
    \begin{subfigure}{0.48\textwidth}
        \centering
        \includegraphics[width=\textwidth]{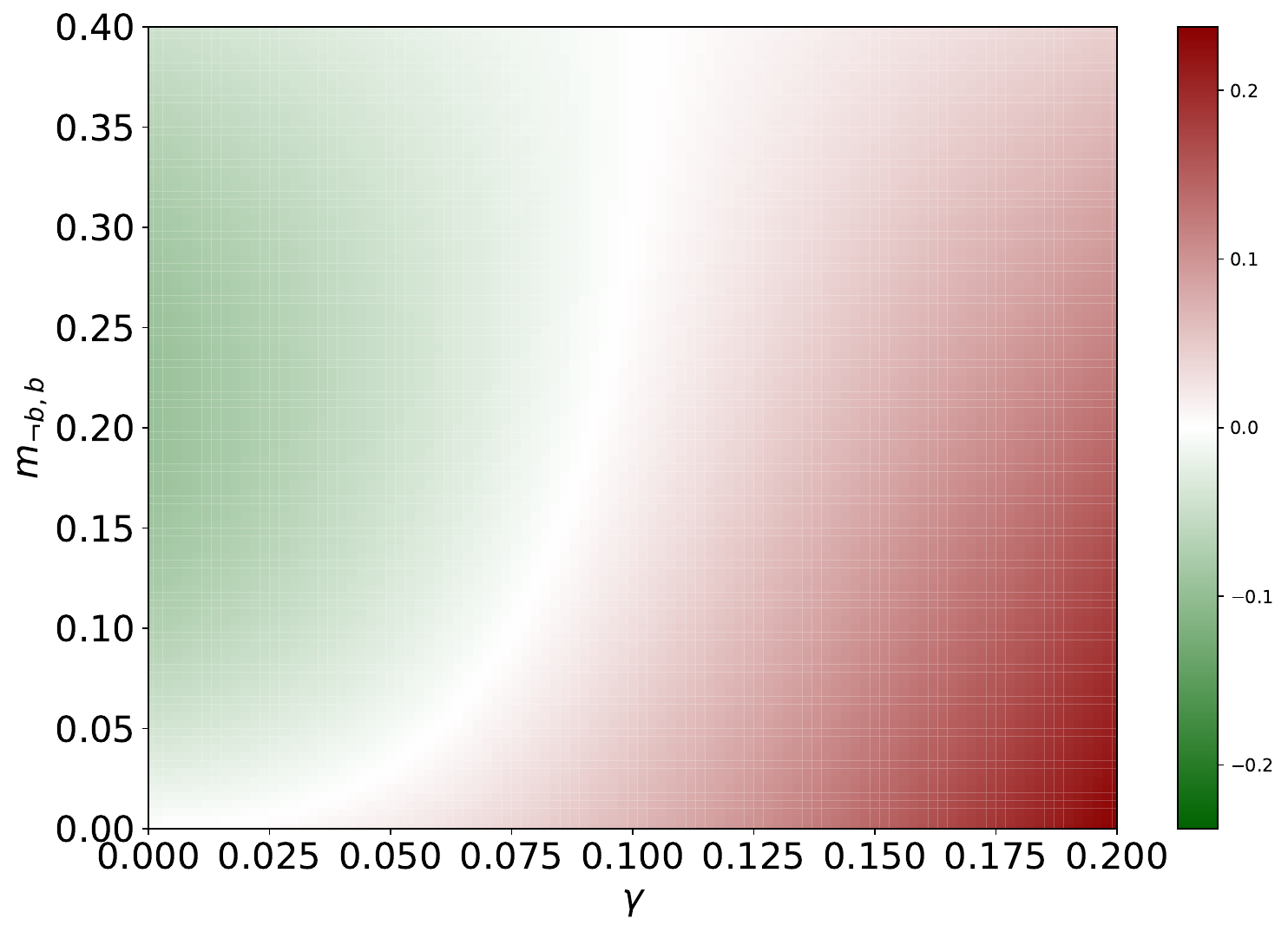} 
        \captionsetup{justification=centering}
        \caption{}
        \label{fig:initial_change}
    \end{subfigure}
    \hfill
    \begin{subfigure}{0.48\textwidth}
        \centering
        \includegraphics[width=\textwidth]{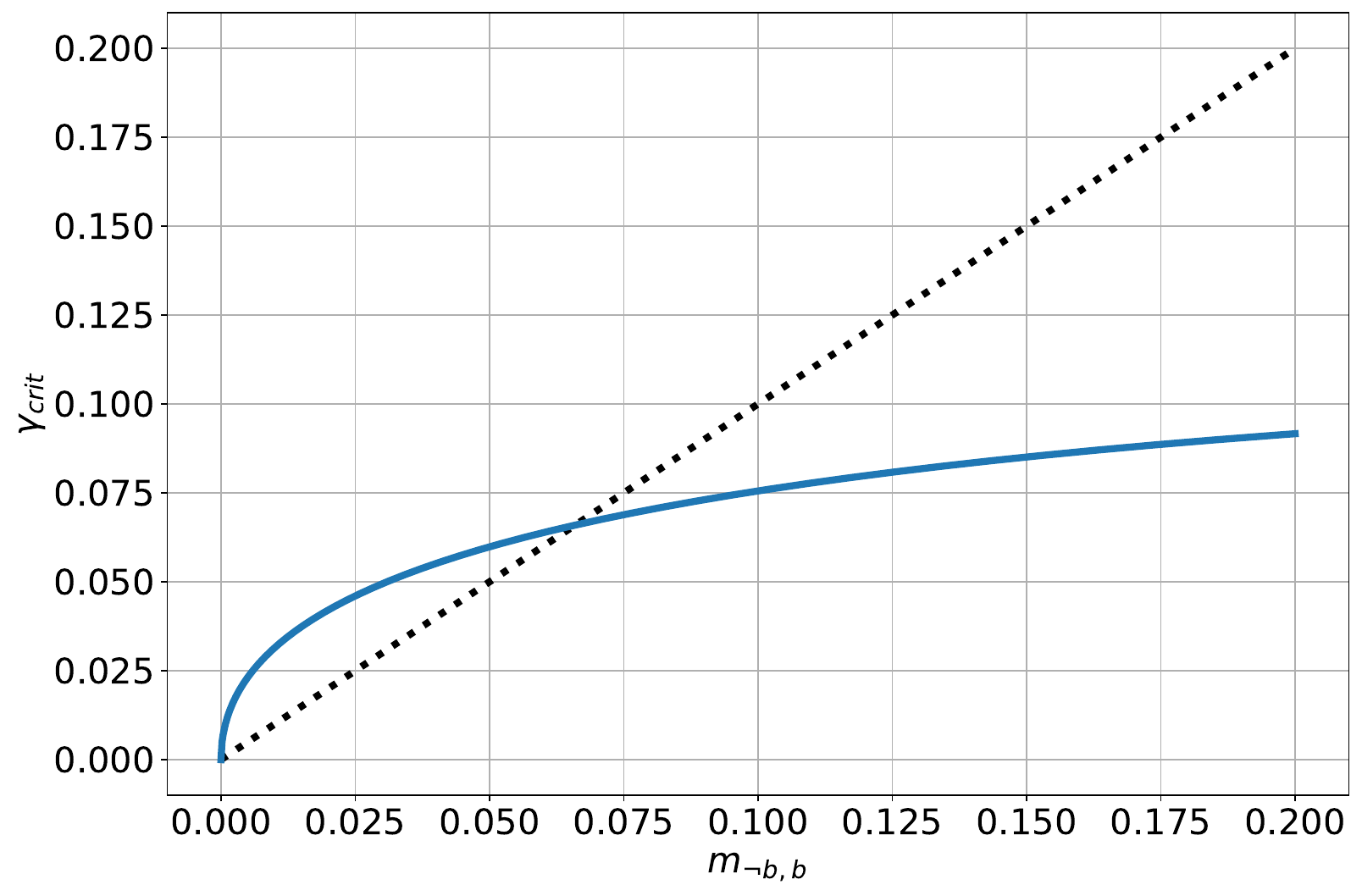} 
        \captionsetup{justification=centering}
        \caption{}
        \label{fig:gamma_crit}
    \end{subfigure}
    \caption{Panel (a): Heatmap of the change in $\eps$ from $N=1$ to $N=3$, $\eps_{N=3}-\eps_{N=1}$, in dependence of the initial error rate $m_{\neg b, b}$ and the $\CNOT$ error $\gamma$. Panel (b): The solid blue lines show the critical threshold $\gamma_{\text{crit}}$ which separates initial improvement $\eps_{N=3}-\eps_{N=1}<0$, from initial worsening $\eps_{N=3}-\eps_{N=1}>0$ in panel (a) as a function of the initial error rate $m_{\neg b, b}$. The dotted black line shows $\gamma_{\text{crit}}=m_{\neg b, b}$.}
\end{figure*}

\begin{figure}
    \centering
    \includegraphics[width=\linewidth]{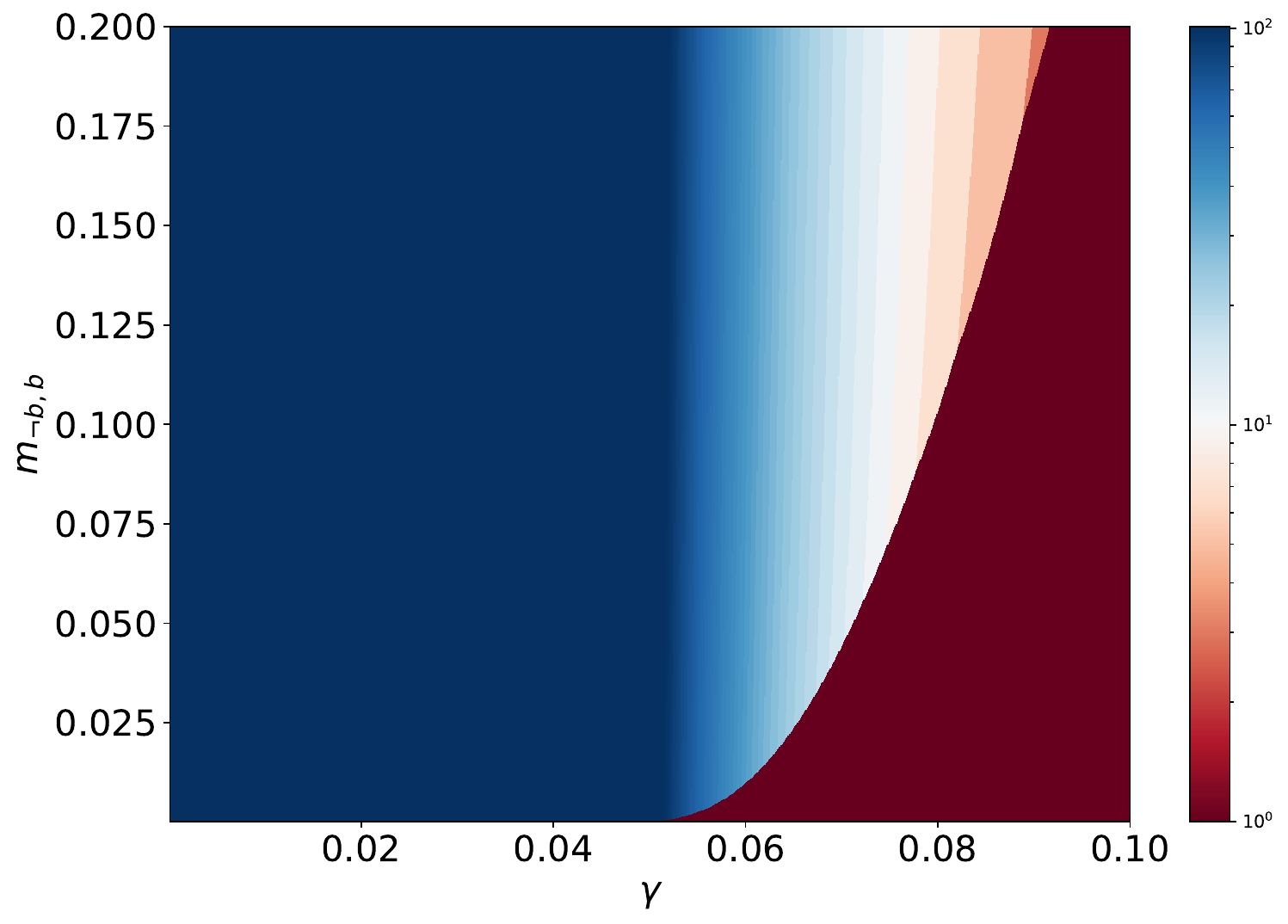}
    \caption{Heatmap of the $N\in [1,101]$ which results in the smallest $\eps$ in dependence of the initial error rate $m_{\neg b, b}$ and the $\CNOT$ error $\gamma$.}
    \label{fig:best_N}
\end{figure}

\begin{figure}
    \centering
    \includegraphics[width=\linewidth]{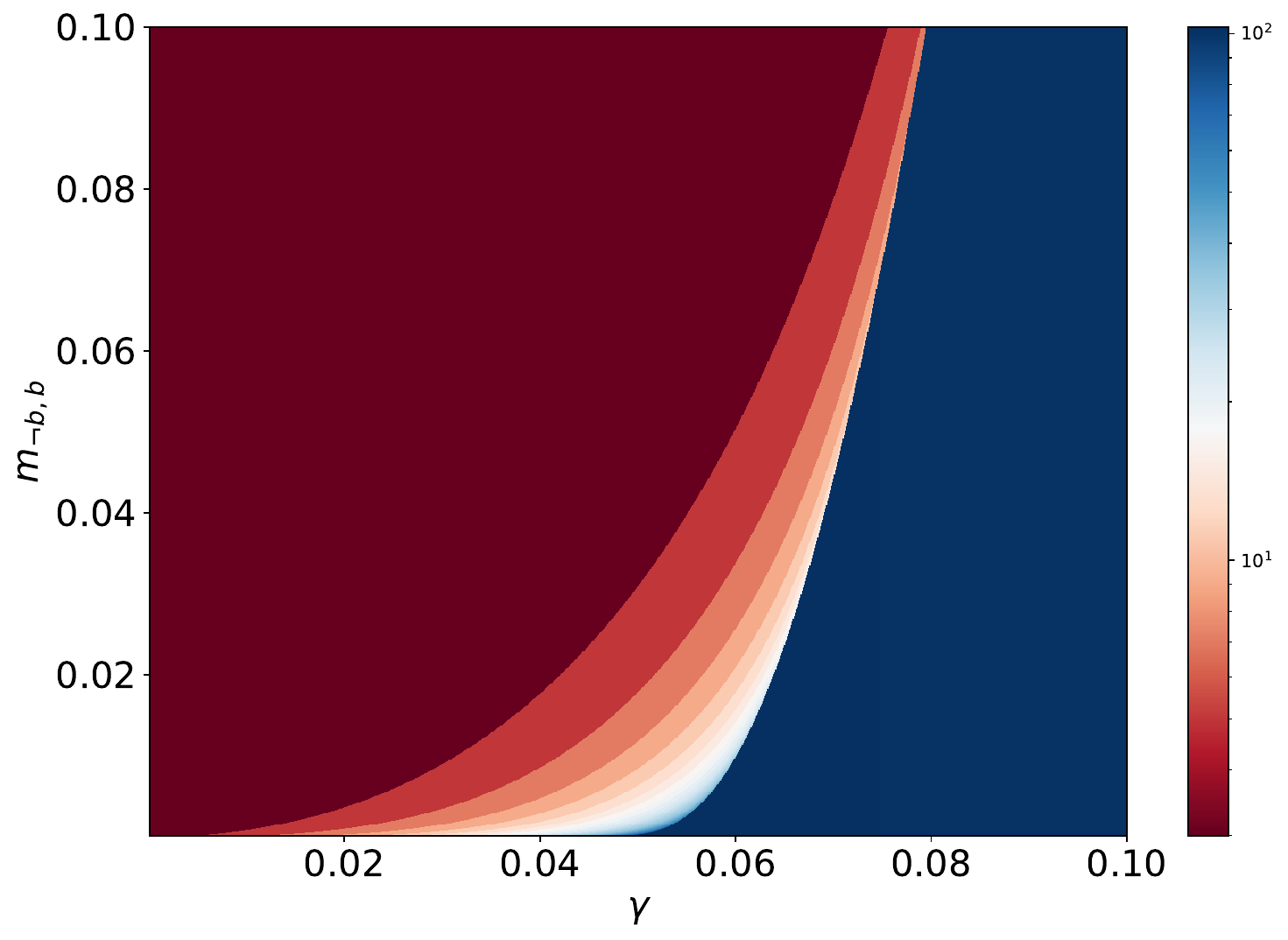}
    \caption{Heatmap of the $N\in[1,101]$ which results in the first $\eps_N<\eps_{N=1}$, in dependence of the initial error rate $m_{\neg b,b}$ and the $\CNOT$ error $\gamma$.
    }
    \label{fig:first_improvement}
\end{figure}

As already discussed in the main text, Fig.~\ref{fig:misident_by_effective_error} and Fig.~\ref{fig:regimes} show three regimes for $\eps_{\text{est}}$ in dependence of the $\CNOT$ error $\gamma$: 
\begin{itemize}
    \item Immediate improvement when we increase $N$
    \item Initial worsening, then improvement when we increase $N$
    \item No improvement at all when we increase $N$
\end{itemize}
In this section, we will discuss the different regimes in more detail.
Fig.~\ref{fig:initial_change} shows the change in the misidentification probability $\eps$ from $N=1$ (no verification register) to $N=3$ in dependence of the initial error rate $m_{\neg b, b}$ and the $\CNOT$ error $\gamma$ (cf. Eq.~\eqref{eq:misident_logV}). In the case that $\gamma$ is too large, $N$ needs to be sufficiently large as we will not see an initial improvement for small $N$. The critical threshold $\gamma_{\text{crit}}$ which separates initial improvement $\eps_{N=3}-\eps_{N=1}<0$, from initial worsening $\eps_{N=3}-\eps_{N=1}>0$ is shown in Fig.~\ref{fig:gamma_crit} as a function of the initial error rate $m_{\neg b, b}$. If the $\CNOT$ error $\gamma$ is less than $\gamma_{\text{crit}}$, we are in the region of immediate improvement and otherwise we get an initial worsening of the misidentification probability.
To answer the question of which $N$ gives the lowest misidentification probability $\eps$ for a given initial error $m_{\neg b, b}$ and a given $\CNOT$ error $\gamma$, we calculated this best $N$ iteratively from Eq.~\eqref{eq:misident_logV} and plotted it as a function of $m_{\neg b, b}$ and $\gamma$ in Fig.~\ref{fig:best_N}. The plot shows that for realistic $\CNOT$ errors $\gamma<5\%$, a bigger N is (within reasonable limits) better once the initial worsening is overcome for any initial errors $m_{\neg b, b}\leq0.2$. If the $\CNOT$ error is higher than $5\%$, there exists a transition region depending on the initial error rate $m_{\neg b, b}$ for which there is an optimal $1<N< 101$ corresponding to the minimal~$\eps$. For errors $\gamma$ and $m_{\neg b, b}$ beyond this transition region, larger $N$ do not yield a smaller $\eps$ compared to $N=1$ and thus yield no improvement. 
Similar to the best $N$ with the smallest misidentification probability $\eps$ in Fig.~\ref{fig:best_N}, we plotted the $N\leq 101$ which gives a first improvement in $\eps$ compared to $N=1$ in Fig.~\ref{fig:first_improvement}. We see that there exists a large region of immediate improvement, a small transition region where $N\sim 10$ is required for improvements and then a fairly immediate jump to no improvements at all if $\gamma$ becomes too large.

\end{appendices}

\end{document}